\documentclass[prl, aps, showpacs,superscriptaddress,twocolumn,longbibliography]{revtex4-2}

\makeatletter
\AtBeginDocument{\let\LS@rot\@undefined}
\makeatother

\usepackage{hyperref}
\hypersetup{
     colorlinks=true,
     linkcolor=magenta,
     filecolor=blue,
     citecolor=blue,      
     urlcolor=cyan,
     }
\usepackage{color}
\usepackage[usenames,dvipsnames]{xcolor}
\usepackage{amsmath,amsthm,amssymb}
\usepackage{graphicx} 
\usepackage{float}
\usepackage{epsfig}
\usepackage{bm}
\usepackage{mathrsfs}
\usepackage{multirow}
\usepackage[all]{xy}
\usepackage{pbox}
\usepackage{verbatim}
\usepackage{braket}
\usepackage{mathtools}
\usepackage{bm}
\usepackage{tikz}
\usepackage{xcolor}
 
\usepackage{mathtools}

\usepackage{mathtools}
\usepackage{tabstackengine}
\usepackage{enumerate}   
\usepackage[normalem]{ulem}
\stackMath
\DeclareMathOperator{\Tr}{Tr}

\newcommand{\er}[1]{Eq.~\eqref{#1}}
\newcommand{\ers}[2]{Eqs.~(\ref{#1}-\ref{#2})}

\newcommand{\Er}[1]{Equation~\eqref{#1}}

\newcommand{\beq}{\begin{equation}}
\newcommand{\eeq}{\end{equation}}

\begin{document}  

\title{Dynamical heterogeneity and large deviations in the open quantum East glass model from tensor networks}

\author{Luke Causer}
\affiliation{School of Physics and Astronomy, University of Nottingham, Nottingham, NG7 2RD, UK}
\affiliation{Centre for the Mathematics and Theoretical Physics of Quantum Non-Equilibrium Systems,
University of Nottingham, Nottingham, NG7 2RD, UK}
\author{Mari Carmen Ba\~nuls}
\affiliation{Max-Planck-Institut f\"ur Quantenoptik, Hans-Kopfermann-Str.\ 1, D-85748 Garching, Germany}
\affiliation{Munich Center for Quantum Science and Technology (MCQST), Schellingstr.\ 4, D-80799 M\"unchen, Germany}
\author{Juan P. Garrahan}
\affiliation{School of Physics and Astronomy, University of Nottingham, Nottingham, NG7 2RD, UK}
\affiliation{Centre for the Mathematics and Theoretical Physics of Quantum Non-Equilibrium Systems,
University of Nottingham, Nottingham, NG7 2RD, UK}

\begin{abstract}
    We study the non-equilibrium dynamics of the dissipative quantum East model via numerical tensor networks. We use matrix product states to represent evolution under quantum-jump unravellings for sizes beyond those accessible to exact diagonalisation. This allows us to demonstrate that dynamical heterogeneity accompanies slow relaxation, in analogy with what is seen in classical glassy systems. Furthermore, using variational matrix product operators we: (i) 
    compute the spectral gap of the Lindbladian, and show that glassiness is enhanced in the presence of weak quantum fluctuations compared to the pure classical case, and
    (ii) obtain the dynamical large deviations by calculating the leading eigenvector of the tilted Lindbladian, and find 
    clear evidence for a first-order active-inactive dynamical phase transition. We also show how to directly sample the rare quantum trajectories associated to the large deviations.
\end{abstract}

\maketitle

\textbf{\em Introduction.-} 
Kinetically constrained models (KCMs) 
serve as an important paradigm for understanding non-equilibrium dynamics.
Originally introduced to model the steric interactions responsible for the slow relaxation of structural glasses 
\cite{fredrickson1984kinetic,palmer1984models,jackle1991a-hierarchically,kob1993kinetic}, 
classical KCMs provide the combination of simple static properties with complex cooperative dynamics due to constraints
\cite{ritort2003glassy,chandler2010dynamics,garrahan2011kinetically,speck2019dynamic,hasyim2023emergent}. Quantum KCMs in turn appear in several contexts, one being Rydberg atoms where strong interactions are responsible for either ``Rydberg blockade'' (encoded in the PXP model \cite{fendley2004competing,lesanovsky2011many-body,turner2018weak}) 
or ``facilitated'' dynamics 
\cite{ates2007antiblockade,amthor2010evidence,lesanovsky2014out-of-equilibrium,hoening2014antiferromagnetic,valado2016experimental,ostmann2019localization,causer2020dynamics}; another, a scenario for non-ergodicity due to constraints rather than disorder
\cite{horssen2015dynamics,lan2018quantum,pancotti2020quantum,deger2022arresting,deger2022constrained,
valencia-tortora2022kinetically,carollo2022signatures,zadnik2021the-folded,bertini2023localised,bertini2023exact}.

In this paper, we focus on the dynamics of quantum KCMs in the presence of an environment, specifically on the open quantum East model (OQEM) \cite{olmos2012facilitated}, which generalises the well-studied East model to a quantum dissipative setting. We consider both the typical dynamics and the large deviations of the OQEM using numerical tensor networks (TNs) \cite{orus2019tensor,banuls2023tensor}. We simulate quantum trajectories of pure states under quantum-jump unravellings for large system sizes using matrix product states (MPS) 
\cite{verstraete2008matrix,schollwock2011the-density-matrix}
as an ansatz for the wavefunction 
\cite{daley2014qtraj}, an approach well suited to this problem, as constrained dynamics limits the growth of entanglement. 
We also use variational matrix product operators (vMPOs) 
\cite{cui2015variational} to directly approximate the eigenvectors of the Lindbladian and: 
(i) estimate the spectral gap of the generator of the dynamics in order to  quantify the relaxation time, showing convincingly ``re-entrant'' behaviour
\cite{markland2011quantum}, whereby a small amount of quantum fluctuations  slows dynamics compared to the classical limit; (ii) calculate the dynamical large deviations (LDs) 
\cite{touchette2009the-large,garrahan2018aspects,jack2020ergodicity}, showing the existence of an active-inactive dynamical phase transition, in analogy with the classical East model \cite{garrahan2007dynamical}. 
Our results extend the applicability of TN methods for the study of rare events in classical stochastic systems \cite{gorissen2009density-matrix,gorissen2012current,gorissen2012exact,banuls2019using,helms2019dynamical,helms2020dynamical,causer2021optimal,causer2022finite,gu2022tensor-network,strand2022using,causer2023optimal} to quantum stochastic systems.

\textbf{\em Open Quantum East Model.-}
We consider an open quantum system whose evolution is given by a Lindblad–Gorini–Kossakowski–Sudarshan 
\cite{lindblad1976on-the-generators,gorini1976completely}
master equation, $\dot{\rho_{t}} = \mathcal{L}[\rho_{t}]$, where $\rho_{t}$ is the density matrix of the system at time $t$, with super-operator $\mathcal{L}$ that generates this dynamics (the so-called Lindbladian) \cite{gardiner2004quantum},
\beq
    \mathcal{L}[\boldsymbol{\cdot}] = -i[\hat{H}, \boldsymbol{\cdot}] + \mathcal{D}[\boldsymbol{\cdot}].
    \label{lindblad}
\eeq
The OQEM \cite{olmos2012facilitated,rose2022hierarchical} is defined in terms of a one-dimensional lattice of $N$ qubits with Hamiltonian
\beq
    \hat{H} = \Omega \sum_{j=1}^{N} \hat{f}_{j}\hat{\sigma}_{j}^{x},
    \label{H}
\eeq
and dissipator
\beq
    \mathcal{D}[\boldsymbol{\cdot}] = \sum_{\alpha = +, -}\sum_{j=1}^{N} \hat{J}_{\alpha, j}\boldsymbol{\cdot} \hat{J}_{\alpha, j}^{\dagger} -\frac{1}{2}\{ \hat{J}_{\alpha, j}^{\dagger}\hat{J}_{\alpha, j}, \boldsymbol{\cdot}\}
    \label{D}
\eeq
with jump operators 
\beq
    \hat{J}_{+, j} = \sqrt{\gamma} \hat{f}_{j} \hat{\sigma}^{+}_{j}, \,\, \hat{J}_{-, j} = \sqrt{\kappa} \hat{f}_{j} \hat{\sigma}^{-}_{j} ,
    \label{J}
\eeq
where $\hat{\sigma}^{+}_{j} = \ket{1}\bra{0}$ and $\hat{\sigma}^{-}_{j} = \ket{0}\bra{1}$ are the spin-$1/2$ ladder operators acting on site $j$, with $\hat{\sigma}^{x}_{j} = \hat{\sigma}^{+}_{j} + \hat{\sigma}^{-}_{j}$. 

\begin{figure}[t]
    \centering
    \includegraphics[width=\linewidth]{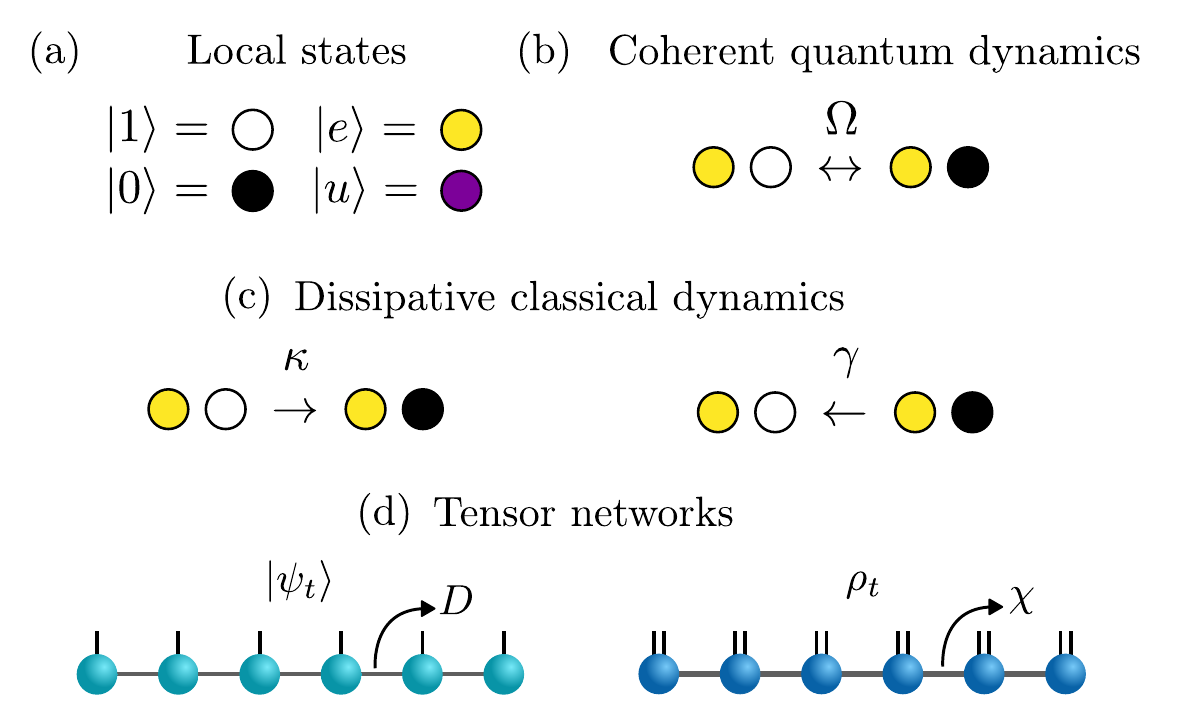}
    \caption{\textbf{Open quantum East model and TNs.} 
    (a) Local classical (or computational) basis states $\ket{1}$  and $\ket{0}$ (``up'' or ``down'') are represented by 
    empty (white) or filled (black) circles,
    respectively.
    The local ``excited'' and ``unexcited'' states, $\ket{\rm e}$ and $\ket{\rm u}$, are superpositions (see Refs.~\cite{olmos2012facilitated,rose2022hierarchical} for their specific forms), 
    shown as bright (yellow) and dark (purple) circles, respectively.
    (b) Coherent transitions are allowed only if the neighbour to the left has a projection onto the excited state, see \er{H}. 
    (c) Same for dissipative transitions, see \er{J}. 
    (d) MPS are used to approximate wavefunctions, $\ket{\psi_{t}}$, and MPOs density matrices, $\rho_{t}$. The internal bonds have dimensions $D$ and $\chi$, respectively. Open (physical) legs have dimensions $2$ and $2 \times 2$, respectively.
    }
    \label{fig: model_tns}
\end{figure}

Both the term of the Hamiltonian \eqref{H} and the jump operators \eqref{J} that act on site $j$ depend on the {\em kinetic constraint} operator $\hat{f}_{j}$ \cite{olmos2012facilitated}. This is a projector that depends on the state of the neighbouring site $j-1$ and acts as an operator-valued rate that determines whether a local transition can take place depending on its neighbours, in analogy with how classical KCMs are defined \cite{ritort2003glassy}. The form of $\hat{f}_{j}$ is chosen \cite{olmos2012facilitated} such that the stationary state of \er{lindblad} is the same as in the unconstrained case (when all $\hat{f}_{j} = 1$), 
\beq
    \rho^{\rm ss} = \bigotimes_{j=1}^{N} 
    \left( p_{\rm e}\ket{e}\bra{e}_{j} + p_{\rm u}\ket{u}\bra{u}_{j} \right), 
    \label{ss}
\eeq
where $\ket{u}_{j}, \ket{e}_{j}$ are the eigenstates of the local stationary density matrix at $j$, and $p_{\rm u}, p_{\rm e}$ its eigenvalues \cite{olmos2012facilitated}, with $p_{\rm u} + p_{\rm e} = 1$. 
We label the states according to their probabilities, such that $p_{\rm e} \leq p_{\rm u}$, and we call 
$\ket{e}_{j}$ {\em excited} and $\ket{u}_{j}$ local {\em unexcited} states. 
The constraint then reads \cite{olmos2012facilitated}
\beq
    \hat{f}_{j} = \ket{e}\bra{e}_{j-1} \equiv \hat{P}_{j-1}^{e},
    \label{constraint}
\eeq
where $\hat{P}_j^{e}$ is the projector onto the local excited state. The local states and transitions of the OQEM are illustrated in Fig.~\ref{fig: model_tns}(a-c). As shown originally in Ref.~\cite{olmos2012facilitated}, the choice \er{constraint} guarantees that the stationary state of the OQEM is the product state \er{ss}. In the limit $\Omega = 0$ the dynamics generated by \ers{lindblad}{J} is equivalent to that of the classical East model at temperature $1/\ln{(\gamma / \kappa)}$ [cf.\ Fig.~\ref{fig: model_tns}(c)], so that by tuning $\Omega$ we can investigate the interplay of classical and quantum relaxation mechanisms. For convenience, in what follows we study the OQEM with the open boundary condition $\hat{P}_{0}^{e}= 1$.

\textbf{\em Tensor networks.-}
TNs are efficient parametrisations of high dimensional objects, such as states and operators, in terms of smaller tensorial 
objects~\cite{verstraete2008matrix,schollwock2011the-density-matrix,orus2014a-practical,bridgeman2017hand-waving,silvi2019the-tensor,ran2020tensor,okunishi2022developments,banuls2023tensor}.
An especially useful decomposition
for the wavefunction of a one dimensional system is the {\em matrix product state} (MPS) \cite{schollwock2011the-density-matrix}, shown pictorially in Fig.~\ref{fig: model_tns}(d): each lattice site is given its own rank-3 tensor (blue circles), with {\em physical} dimension $d$ ($=2$ in our case) for the vertical legs, and virtual or {\em bond} 
dimension $D$ for each of the horizontal legs.
Each tensor is connected to its neighbouring tensors along the lattice sites via its virtual legs, with the  many-body state obtained by contracting (i.e. multiplying and summing over the connected indices)
all virtual legs. The number of parameters in such an MPS is $\mathcal{O}(NdD^{2})$, which is much smaller than what is required to represent a generic vector for large $N$, at the price of only accurately describing 
a subset of the Hilbert space with entanglement upper-bounded by $S_{E} \leq 2 \log D$, which happens to approximate well ground states of local Hamiltonians~\cite{cirac2021matrix}.

A similar ansatz can be applied to operators acting on a 1D system. Such a {\em matrix product operator} (MPO)~\cite{verstraete2004matrix,zwolak2004mixed-state,pirvu2010matrix}
is depicted in Fig.~\ref{fig: model_tns}(d) for the density matrix: each local tensor is of rank-4, with two physical dimensions (each $d=2$) and two virtual dimensions (each $\chi$).
For physical states, this MPO needs to be positive, and while, strictly speaking, positivity cannot even be decided at the level of the local tensors \cite{kliesch2014matrix-product,cuevas2016lim}, in practice 
many MPO algorithms can approximately preserve it
\footnote{An ansatz worth mentioning is the {\em purification ansatz}~\cite{verstraete2004matrix,cuevas2013purifications,kliesch2014matrix-product}.While this preserves positivity, it is computationally more expensive and more restricted.}.

We use both MPS and MPOs to investigate the OQEM for sizes beyond those accessible to exact diagonalisation:
(i) MPS to simulate the quantum trajectories of pure states evolving under a quantum jump unravelling \cite{finsterholzl2020using};
(ii) variational MPOs (vMPOs) to estimate the spectral gap of the Lindbladian, and to estimate the leading eigenvectors of {\em tilted} Lindbladians that encode the large deviations \cite{garrahan2010thermodynamics}; 
and (iii) both in conjunction to efficiently generate quantum trajectories that realise the dynamical large deviations. For details of the methods used see Ref.~\cite{supplemental}.

\begin{figure}[t] 
    \centering
    \includegraphics[width=0.9\linewidth]{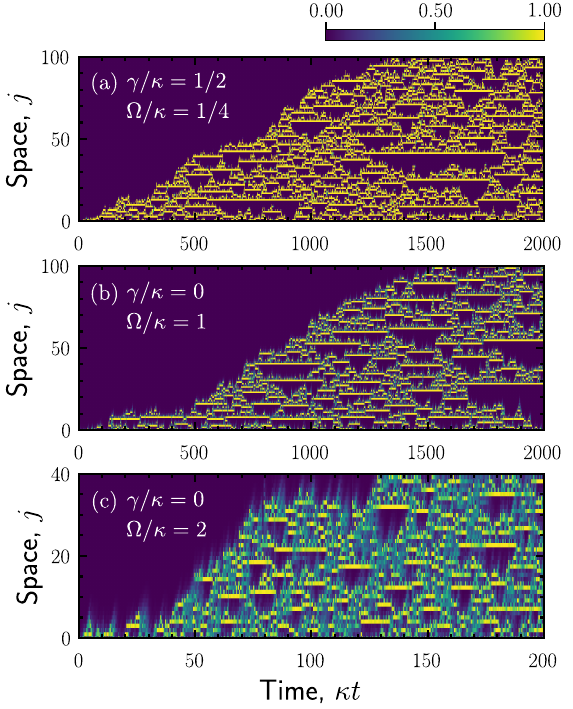}
    \caption{\textbf{Quantum jump trajectories of the OQEM.} 
    (a) Local projection onto the excited, $\braket{\hat{P}_{j}^{e}}_{t}$, as a function of rescaled time, $\kappa t$, from a quantum trajectory starting from 
    $\ket{\psi_{0}} = \otimes \bigotimes_{j=1}^{N} \ket{u}_{j}$ , for $\gamma/\kappa=1/2$, $\Omega/\kappa=1/4$. (Note that the open boundary conditions are such that site $j=1$ is always allowed to flip.) 
    (b) Same for $\gamma=0$, $\Omega/\kappa=1$.
    (c) Same for $\gamma=0$, $\Omega/\kappa=2$ (note that both system size and overall time are smaller than in other panels).
    }
    \label{fig: trajectories}
\end{figure}

\textbf{\em Quantum trajectories and dynamical heterogeneity.-}
We study first the quantum trajectories of the OQEM under a quantum jump unravelling \cite{plenio1998the-quantum-jump}. This corresponds to a continuous-time quantum Markov chain where the pure state $\ket{\psi_{t}}$ is evolved deterministically, 
$\ket{\psi_{t+\Delta t}} = e^{-i\Delta t \hat H_{\rm eff}} \ket{\psi_{t}}/||e^{-i\Delta t \hat H_{\rm eff}} \ket{\psi_{t}}||$ 
(with $\hat{H}_{\rm eff} = \hat{H} - \frac{i}{2}\sum_{\alpha, j} \hat{J}^{\dagger}_{\alpha, j}\hat{J}_{\alpha, j}$),  punctuated stochastically by quantum jumps,
$\ket{\psi_{t}} \to \ket{\psi'_{t}} = \hat{J}_{\mu} \ket{\psi_{t}}/|| \hat{J}_{\mu} \ket{\psi_{t}}||$, 
with transition rates 
$w_{\mu}(\ket{\psi_{t}}) = {\braket{\psi_{t} | \hat{J}^{\dagger}_{\mu} \hat{J}_{\mu} | \psi_{t}}}{\braket{\psi_{t} | \psi_{t}}}$.
By modelling the state using a MPS ansatz with a bond dimension that can be dynamically adjusted \cite{finsterholzl2020using} to account for the fluctuating levels of entanglement, we can simulate these trajectories for sizes much larger than those accessible by exact methods in previous works~\cite{olmos2012facilitated,rose2022hierarchical}. For implementation details see Ref.~\cite{supplemental}.


Figure \ref{fig: trajectories} shows trajectories starting with all sites in the unexcited state, $\ket{\psi_{0}} = \bigotimes_{j=1}^{N} \ket{u}_{j}$, and we 
plot the local projections onto the excited state, $\braket{\hat{P}_{j}^{e}}_{t} = \braket{\psi_{t} | \hat{P}_{j}^{e} | \psi_{t}}$.
Panel (a) shows a trajectory for $\gamma / \kappa = 1/2$ and $\Omega / \kappa = 1/4$ with system size $N = 100$ and total time $\kappa T = 2000$: despite the coherent driving, it being relatively weak means that quantum superpositions are mostly suppressed; the trajectory 
looks like those in the classical model, 
with the characteristic ``space-time bubbles'' of inactivity that give rise to dynamical heterogeneity \cite{garrahan2002geometrical}.
Panel (b) shows a trajectory for $\gamma / \kappa = 0$ (i.e.\ ``zero temperature'') and $\Omega / \kappa = 1$: the absence of the dissipative excitation process means that the coherent driving is the only mechanism to give rise to relaxation; the dynamical fluctuations here are greater than in the weakly perturbed classical case, and the system exhibits a large degree of dynamical heterogeneity that is quantum in origin.
Panel (c) shows the same but for lager driving, $\Omega / \kappa = 2$ (for $N = 40$ and $T = 100$ as relaxation is faster in this case): the quantum effects are more pronounced and the increased coherent driving reduces dynamical fluctuations.

\textbf{\em Relaxation timescale and
re-entrant glassy behaviour.-} 
The quantum trajectories of Fig.~\ref{fig: trajectories} illustrate the basic physics of the OQEM. Due to the constraint, relaxation propagates from regions with excitations to regions without. For an empty initial condition such as that of Fig.~\ref{fig: trajectories} (with an active boundary), in the classical limit it is known that the excited region expands ballistically into the unexcited region,  eventually relaxing the whole system (proven rigorously in the asymptotic limit in Ref.~\cite{blondel2013front} for such KCMs). The finite space and time trajectories of Fig.~\ref{fig: trajectories} suggest that this is also the case for the OQEM. Relaxation speed is determined both by the dissipative excitation process, controlled by the ``classical'' rate $\gamma / \kappa$, and by the coherent process, of scaled strength $\Omega / \kappa$. As in the classical case \cite{merolle2005space-time,garrahan2007dynamical,katira2016pre-transition,klobas2023exact}, the trajectories show coexistence of space-time regions of high and low activity. For small $\gamma / \kappa$ and $\Omega / \kappa$, relaxation is slow and glassy. 

\begin{figure}[t]
    \centering
    \includegraphics[width=0.9\linewidth]{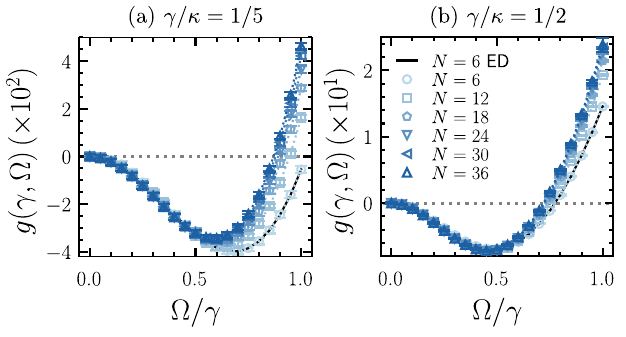}
    \caption{\textbf{Re-entrant glassiness due to quantum fluctuations.} 
    (a)
    Relative gap $g(\gamma, \Omega)$ as a function of coherent driving relative to dissipative excitation rate, $\Omega / \gamma$,  for $\gamma / \kappa = 1/5$. 
    The dotted line at zero separates the regime of faster-than-classical relaxation ($g>0$) from that of slower-than-classical ($g<0$). 
    The symbols show the vMPO results for sizes $N=6-36$, and the dashed line shows the exact diagonalization result for $N=6$ for reference.
    (b) Same for $\gamma / \kappa = 1/2$. 
    }
    \label{fig: gaps}
\end{figure}

The typical relaxation timescale from an arbitrary configuration 
\footnote{This is different from the timescale to relax from the most unfavourable initial state, as in Fig.~\ref{fig: trajectories}, whose timescale is related to the ``cutoff phenomenon'' \cite{aldous1986shuffling} of Markov chains.
}
is given by the inverse of the spectral gap of the Lindbladian \cite{cai2013algebraic,znidaric2015relaxation,macieszczak2016towards,zhou2022exponential}. We can estimate the gap using a variational optimisation of the eigenvectors of the Lindbladian represented as a MPO ~\cite{cui2015variational} (see Ref.~\cite{supplemental} for details). The results of this approach are shown in Fig.~\ref{fig: gaps} where we plot the gap of the OQEM relative to the classical case,
\beq
    g(\gamma / \kappa, \Omega / \kappa) = \frac{\delta(\gamma / \kappa, \Omega / \kappa)}{\delta(\gamma / \kappa, 0)} - 1,
    \label{g}
\eeq
as a function of the coherent driving $\Omega / \kappa$ for two values of the dissipative excitation rate, panel (a) showing $\gamma / \kappa = 1/5$, and panel (b) $\gamma / \kappa = 1/2$. The definition \er{g} makes $g>0$ if relaxation is faster than in the classical limit, and $g<0$ if it is slower. Figure~3 shows that the addition of a small amount of quantum coherence ($\Omega / \kappa \gtrsim 0$) enhances glassiness within the dynamics, a phenomenon sometimes referred to as {\em quantum re-entrance} \cite{markland2011quantum}. When the strength of the quantum coherence is increased enough, the gap eventually increases resulting in a faster than classical relaxation. Our results seem to confirm the trend observed in Ref.~\cite{olmos2012facilitated} for a different KCM - the open quantum Fredrickson-Andersen model - for sizes up to $N = 5$ 
(such that the Lindbladian can be diagonalised numerically exactly). In contrast, with our TN methods here we are able to reach up to sizes $N = 36$ for the OQEM. 
Note that in Fig.~\ref{fig: gaps} the relative gap \er{g} appears to converge with $N$, which suggests that the observed behaviour holds for larger $N$. Furthermore, the error bars for each data point are small (see Ref.~\cite{supplemental} for details), indicating a reasonable approximation of eigenvalues of $\mathcal{L}$: while there is no guarantee that the TN method is approximating the eigenvalue with the smallest real part, at worst our results are an estimate of an upper bound to the spectral gap.

\begin{figure*}[t]
    \centering
    \includegraphics[width=\linewidth]{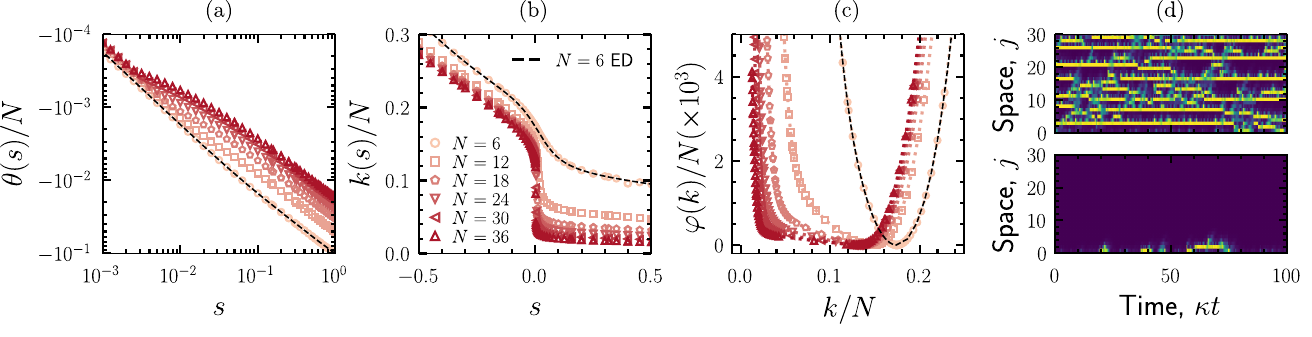}
    \caption{\textbf{Dynamical large deviations and optimal sampling.} 
    (a) SCGF $\theta(s) / N$ as a function of $s$ from the vMPOs approximation to the leading eigenmatrix $L_s$. We show sizes $N = 6 - 36$ (symbols) together with $N = 6$ (dashed line) from exact diagonalisation for comparison. At small $s$, the SCGF goes as $\theta(s) \approx -s k(0)$  (dashed line), where $k(0)$ is the average dynamical activity at stationarity per unit time. For larger $s$ it departs from the linear response behaviour, with the crossover taking place at smaller $s$ with increasing size, indicative of a phase transition. 
    (b) Dynamical activity, $k(s) / N = -\theta'(s) / N$. The sharp drop (which gets more pronounced with size) indicates a first-order transition.
    (c) Rate function $\varphi(k) / N$.  The broadening corresponds to (active/inactive) phase coexistence, which gets sharper with increasing size. (d) Characteristic quantum trajectories of the active ($s=-0.1$, top panel) and inactive ($s=0.1$, bottom panel) phases sampled using the optimal (Doob) dynamics. 
    }
    \label{fig: lds}
\end{figure*}

\textbf{\em Large deviations and dynamical phase transitions.-}
The pronounced fluctuations in the quantum trajectories of the OQEM shown in Fig.~\ref{fig: trajectories} are suggestive of coexistence between dynamics that is active and fast relaxing and dynamics that is inactive and unable to relax. Classically \cite{merolle2005space-time,garrahan2007dynamical} this behaviour is most naturally quantified through the statistics of the {\em dynamical activity} \cite{lecomte2007a-numerical,garrahan2018aspects,maes2020frenesy:}. The equivalent for open quantum systems \cite{garrahan2010thermodynamics} is through the statistics of the total number of quantum jumps. 

The probability to observe a number $K$ of quantum jumps in a quantum trajectory of time extent $t$ is $P_{t}(K)$. For large $t$ this probability will obey a large deviation (LD) principle \cite{touchette2009the-large}, $P_{t}(K) \asymp e^{-t\varphi(K/t)}$, with $\varphi(k)$ the LD {\em rate function}
\footnote{
    For the OQEM this is guaranteed as ${\mathcal L}$ is gapped and its stationary state is unique.
}. The same statistics is encoded in the moment generating function $Z_{t}(s) = \sum_{K} P_{t}(K)e^{-sK}$, where $s$ is the {\em counting field} conjugate to $K$ \cite{garrahan2010thermodynamics}. The corresponding LD principle reads $Z_{t}(s) \asymp e^{t\theta(s)}$, where $\theta(s)$ is the scaled cumulant generating function (SCGF) \cite{garrahan2010thermodynamics}.
For open quantum dynamics, the SCGF can be obtained from the {\em tilted} Lindbladian 
$\mathcal{L}_s[\boldsymbol{\cdot}] = -i[\hat{H}, \boldsymbol{\cdot}] + \mathcal{D}_s[\boldsymbol{\cdot}]$ \cite{garrahan2010thermodynamics}, which for the OQEM has dissipator, cf.\ \eqref{D}, 
\beq
    \mathcal{D}_s[\boldsymbol{\cdot}] = \sum_{\alpha = +, -}\sum_{j=1}^{N} 
    e^{-s}
    \hat{J}_{\alpha, j}\boldsymbol{\cdot} \hat{J}_{\alpha, j}^{\dagger} -\frac{1}{2}\{ \hat{J}_{\alpha, j}^{\dagger}\hat{J}_{\alpha, j}, \boldsymbol{\cdot}\} .
    \label{Ds}
\eeq
Specifically, the SCGF $\theta(s)$ is the largest eigenvalue of $\mathcal{L}_s$, with right and left eigenmatrices, $\mathcal{L}_{s}[R_{s}] = \theta(s)R_{s}$ and $\mathcal{L}^{\dagger}_{s}[L_{s}] = \theta(s) L_{s}$ \cite{garrahan2010thermodynamics}.
Knowing $\theta(s)$ we can recover the rate function, and thus the distribution of $K$, by inverting the Legendre transform, $\theta(s) = -\min_{k}[ks + \varphi(k)]$ \cite{touchette2009the-large}.

We can compute the SCGF numerically for the OQEM by estimating either of the eigenmatrices $R_{s}$ or $L_{s}$ using non-Hermitian vMPOs \cite{cui2015variational}. In practice, we find the method is more stable and has a better accuracy when targeting $L_{s}$. The results for the SCGF obtained in this way are shown in Fig.~\ref{fig: lds} for the case of $\gamma / \kappa = 0$ and $\Omega / \kappa = 1$. For these parameters, we are able to find results with our desired accuracy of $\varepsilon \leq 10^{-3}$
for system sizes up to $N = 36$
(see Ref.~\cite{supplemental} for details).
The SCGF $\theta(s) / N$ is shown in Fig.~\ref{fig: lds}(a) as a function of $s$: for small $s \gtrsim 0$, the SCGF follows the (linear response) branch $\theta(s) \approx -sk(0)$ where $k(0)$ is the average $K$ (per unit time) in the stationary state \eqref{ss} (black dashed line), while at $s_{c}(N) \gtrsim 0$ (which decreases with system size) there is a sharp crossover away from linear response. 
The tilted dynamical activity is given by the derivative of the SCGF, $k(s) = -\theta'(s)$, and is shown in Fig.~\ref{fig: lds}(b): this shows that the change at $s_{c}(N) \gtrsim 0$ in the SCGF corresponds to a sharp drop in activity, indicating a first-order dynamical phase transition in the large size limit. Figure~\ref{fig: lds}(c)
shows the corresponding rate function: 
the first-order transition manifests as large fluctuations in $K$ due to the coexistence between an active (large $K$) dynamical phase and and inactive (small $K$) one. 

\textbf{\em Optimal dynamics for sampling the large deviations.-}
Our MPO approximation to the left eigenmatrix of ${\mathcal L}_s$ allows us to realise the optimal dynamics that samples the rare quantum trajectories that realise the large deviations as controlled by $s$. This is done via the quantum generalisation 
\cite{garrahan2010thermodynamics,carollo2018making,carollo2019unraveling,carollo2021large} of the classical generalised {\em Doob transform} \cite{borkar2003peformance,jack2010large,chetrite2015nonequilibrium,garrahan2016classical}. 

For quantum open systems it is important to note that the Lindbladian {\em is not} the quantum equivalent of a classical 
Markov generator, as it only generates the dynamics of the average state. The equivalent operator is what in Refs.~\cite{carollo2019unraveling,carollo2021large} is called ``unravelled generator'', which generates the stochastic quantum trajectories. While one can do a ``quantum Doob transform'' \cite{garrahan2010thermodynamics,carollo2018making} at the level of the Lindbladian and obtain  average behaviour compatible with $s \neq 0$, in order to generate actual rare tajectories at $s \neq 0$ optimally we need to perform a Doob transform at the level of the unravelled generator \cite{carollo2019unraveling,carollo2021large}. This gives a stochastic jump dynamics with the same Hamiltonian, \er{H}, but jumps executed by operators similar in form to those in \er{J} but with modified rates \cite{carollo2021large}
\beq
    w^{s}_{\mu}(\ket{\psi_{t}}) = e^{-s}\frac{\braket{\psi_{t} | \hat{J}^{\dagger}_{\mu} L_{s} \hat{J}_{\mu} | \psi_{t}}}{\braket{\psi_{t} | L_{s} | \psi_{t}}} ,
    \label{doob_rates}
\eeq
where $\mu$ stands for $\alpha=\pm 1, j$ for the OQEM, cf.\ \er{J}. The MPO approximation of $L_{s}$ allows us to efficiently obtain \er{doob_rates} as a TN calculation (see Ref.~\cite{supplemental} for details) in a way that generalises to  quantum stochastic dynamics the TN approach of 
Refs.~\cite{causer2021optimal,causer2023optimal}. 

We show quantum trajectories sampled optimally (i.e.\ on demand at the required value of $s$, without the need for post-processing) using our implementation of the optimal Doob dynamics in Fig.~\ref{fig: lds}(d) for $N = 30$. The top panel is for the active phase ($s = -0.1$) and shows a characteristic trajectory of high activity (without inactive space-time bubbles, cf.\ Fig.~\ref{fig: trajectories}). The bottom panel is for the inactive phase ($s=0.1$) and shows a characteristic trajectory with activity only localised near the boundary.

\textbf{\em Discussion.-}
By means of numerical tensor networks, we have investigated the non-equilibrium dynamics of the dissipative quantum East model. We have shown the existence of dynamical heterogeneities in the quantum trajectories, using an MPS approximation to the stochastic pure states. By estimating the spectral gap of the Lindbladian, we were able to show explicitly that the presence of weak quantum fluctuations increases the relaxation time for the model, thus enhancing glassiness and giving rise to re-entrance, i.e., non-monotonic behaviour of the relaxation time as a function of coherent driving. Finally, we investigated the large deviation statistics of the number of quantum jumps, showing that the model exhibits an active-inactive first-order phase transition, very similar to what occurs in classical KCMs. 
We demonstrated that the features above occur also when the only source of relaxation is through coherent processes (i.e., when the classical processes are at zero temperature). Furthermore, we showed how to efficiently sample trajectories corresponding to the dynamical large deviations by constructing an accurate approximation to the optimal Doob dynamics. Our results here are another step in expanding the range of tensor network methods to study stochastic dynamics and rare events \cite{banuls2019using,helms2019dynamical,helms2020dynamical,causer2021optimal,causer2022finite,gu2022tensor-network,strand2022using,causer2023optimal}.

\textbf{\em Acknowledgements.-}
We acknowledge financial support from EPSRC Grant no.\ EP/V031201/1. 
M.C.B.\ acknowledges support from Deutsche Forschungsgemeinschaft (DFG, German Research Foundation) under Germany's Excellence Strategy -- EXC-2111 -- 390814868 and 499180199 (via FOR5522). L.C. was supported through an EPSRC Doctoral Award Prize. We acknowledge access to the University of Nottingham Augusta HPC service.
 
\nocite{hastings2007an-area, verstraete2004matrix, verstraete2006matrix, eisert2010colloquium:, daley2014qtraj,maki2023montecarlo, trotter1959on-the-product, schollwock2011the-density-matrix, Vovk2022entanglement, mascarenhas2015, chan2005density-matrix, Zhang2020skin, helms2019dynamical, cui2015variational, guo2022variational}
\bibliographystyle{apsrev4-2}
%

\newpage
\onecolumngrid
\section{Supplemental Material} 
\setcounter{figure}{0}
\setcounter{equation}{0}
\makeatletter 
\renewcommand{\thefigure}{S\@arabic\c@figure}
\renewcommand{\theequation}{S\@arabic\c@equation}
\makeatother

\subsection{Dissipative quantum kinetically constrained models}

We start with a collection of non-interacting spins whose dynamics are governed by a Lindblad–Gorini–Kossakowski– Sudarshan equation, c.f. Eq.~1 of the main text.
The system has the Hamiltonian 
\beq
     \hat{H} = \Omega\sum_{j=1}^{N} \hat{\sigma}_{j}^{x}
\eeq
and jump operators 
\beq
     \hat{J}_{+, j} = \sqrt{\gamma}\hat{\sigma}_{j}^{+}, \,\, \hat{J}_{-, j} = \sqrt{\kappa}\hat{\sigma}_{j}^{-},
\eeq
i.e., Eqs.~2 and 4 of the main text with $\hat{f}_{j} = 1$.
Since the problem is non-interacting, the stationary state of $\mathcal{L}$ can be written as a product state of local density matrices,
\beq
     \rho^{\rm ss} = \bigotimes_{j=1}^{N} \rho^{\rm loc}.
\eeq
The local stationary state $\rho^{\rm loc}$ is easily determined by solving $\mathcal{L}\rho^{\rm loc} = 0$ for a system size $N = 1$.
It is simple to verify that the solution to this is 
\beq
     \rho^{\rm loc} = \frac{1}{8\Omega^{2} + (\kappa + \gamma)^2}
     \begin{pmatrix}
          4\Omega^{2} + \gamma(\kappa + \gamma) & -2i\Omega(\kappa - \gamma) \\
          2i\Omega(\kappa - \gamma) &  4\Omega^{2} + \kappa(\kappa + \gamma)
     \end{pmatrix}.
\eeq
This can then be diagonalised into an excited state $\ket{e}$ and unexcited state $\ket{u}$ with eigenvalues 
\beq
     p_{e} = \frac{\kappa \cos^{4}\frac{\theta}{2} + \gamma \sin^{4}\frac{\theta}{2}}{(\kappa + \gamma)\left(\cos^{4}\frac{\theta}{2} + \sin^{4}\frac{\theta}{2}\right)},
     \, \, 
     p_{u} = \frac{\gamma \cos^{4}\frac{\theta}{2} + \kappa \sin^{4}\frac{\theta}{2}}{(\kappa + \gamma)\left(\cos^{4}\frac{\theta}{2} + \sin^{4}\frac{\theta}{2}\right)},
\eeq
where the angle $\theta$ is given by 
\beq
     \cos^{2}(\theta) = \frac{(\kappa + \gamma)^{2}}{16\Omega^{2} + (\kappa + \gamma)^{2}}.
\eeq
For $\kappa > \gamma$, it is always the case that $p_{e} \leq p_{u}$.
Note that while closed-form expressions for $\ket{e}$ and $\ket{u}$ are complicated, they are related to $\ket{0}$ and $\ket{1}$ through a unitary transformation, see the supplemental material of Ref.~\cite{olmos2012facilitated} for more details.

We now wish to add the constraint $\hat{f}_{j}$ to $\hat{H}$ and $\hat{J}_{\alpha, j}$, as described in the main text.
For the classical counterpart, adding the kinetic constraints leave the stationary state unaltered, leaving the statics of the problem trivial.
To generalise these models to open quantum dynamics, we require this to still be the case.
In the supplemental material of Ref.~\cite{olmos2012facilitated}, it is explained that for this to be true, the constraint must satisfy the following three conditions: (i) $\hat{f}_{j}$ is a projector, $\hat{f}_{j}=\hat{f}_{j}^{\dagger} = \hat{f}_{j}^{2}$; (ii) $\hat{f}_{j}$ commutes with all operators that act locally on site $j$; (iii) $\hat{f}_{j}$ commutes with $\rho^{\rm loc}$, $[\hat{f}_{j}, \rho^{\rm loc}]=0$.
These conditions restrict the allowed class of operators to be diagonal in the eigenstates of $\rho^{\rm loc}$, i.e., operators which are diagonal in $\ket{e}$ and $\ket{u}$.
The operator which satisfies this condition and most closely resembles that of the classical East model is $\hat{f}_{j} = \ket{e}\bra{e}_{j-1}$, which coincides with the classical constraint for $\Omega = 0$.

\subsection{Tensor networks}

Tensor network (TN) states are a class of ansatzes for
correlated many-body systems.
Typically, the dimensionality of a quantum many-body system grows exponentially in the number of its constituents. 
This means that an exponential number of parameters are required to describe an arbitrary tensorial object (such as a wavefunction or an operator).
Despite this, it is sometimes possible to efficiently approximate, or even exactly represent objects of interest with a number of parameters that is only polynomial in the number of subsystems. 
For example,ground states of local one-dimensional Hamiltonians can be approximated to a given accuracy with tensors whose dimension scales at most polynomially in the system size
\cite{hastings2007an-area,verstraete2006matrix}.

This observation is the essence of TNs.
TNs encode high-dimensional tensorial objects by decomposing them into a network of smaller tensors.
In doing so, we introduce a number of ``virtual dimensions'', which when contracted over (an operation which multiplies and sums over tensors with shared dimensions), yields a tensor with the desired dimensions.
The expressivity of the ansatz is controlled by the size of the virtual dimensions.
By increasing the size of the virtual dimensions, one is able to describe a broader range of states in the full system.

\subsubsection{Matrix product states}

The best understood and, arguably, most useful in practice 
tensor network ansatz is the matrix product state (MPS). 
For a quantum many-body system with $N$ spins (each with dimension $d=2$), a wavefunction can be described as a vector with $d^{N}$ elements,
\beq
     \ket{\psi} = \sum_{j_{1}=1}^{d} \ldots \sum_{j_{N}=1}^{d} \psi_{j_{1}, \dots, j_{N}} \ket{j_{1} \dots j_{N}},
\eeq
where $j_{k}$ is a basis state for the $k$-th subsystem.
An MPS is a vector where each coefficient in the previous expression can be written as 
a product over $N$ matrices,
\beq
     \ket{\psi} = \sum_{j_{1}=1}^{d} \ldots \sum_{j_{N}=1}^{d} \Tr \left[ A^{(1)}_{j_1} \cdots A^{(N)}_{j_N} \right] \ket{j_{1} \dots j_{N}},
     \label{mps}
\eeq
where each $A^{(k)}_{j_{k}}$ is a $D \times D$ matrix. 
Notice that each spin $k$ has $d$ total matrices, and so when $j_{k}$ is unspecified, we can also consider $A^{(k)}_{j_{k}}$ a rank-3 tensor with dimensions $D \times D \times d$. 
The {\em bond dimension} of the MPS, $D$, controls the amount of information that can be shared between the partitions of the system.
In particular, if ones partitions the system into A and B, where A is all spins $k < l$ and B is all spins $k \geq l$, then the bipartite von Neumann entanglement entropy between the two partitions is bounded by $S_{E} \leq 2 \log D$ \cite{eisert2010colloquium:}.

\begin{figure}
     \centering
     \includegraphics[width=\linewidth]{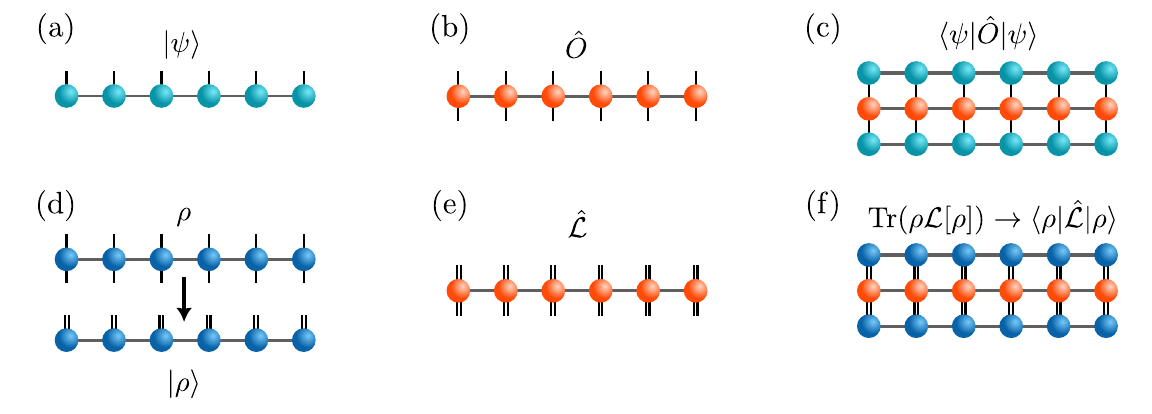}
     \caption{\textbf{Tensor network states.} 
     Diagrammatic representation of TNs.
     Shapes correspond to tensors, and the number of legs gives each tensors rank.
     Closed legs indicate contractions overs tensors.
     (a) The state $\ket{\psi}$ as an MPS.
     (b) The operator $\hat{O}$ as an MPO.
     (c) The inner product $\braket{\psi | \hat{O} | \psi}$ as a TN.
     (d) The density matrix $\rho$ as an MPO. It can be recast as an MPS using the Choi-Jamiołkowski isomorphism.
     (e) The Lindbladian superoperator $\mathcal{L}$ recast as an MPO.
     (f) The expectation of a density matrix with the Lindbladian as a TN.
     }
     \label{fig: mps}
 \end{figure}

As described in the main text, it is often convenient to represent TNs using a diagrammatic notation.
For each tensor in the network, we draw a shape, with a leg for each dimension. 
For the virtual bond dimensions, this is a leg that connects to another tensor.
For the physical dimensions, this could could be left open, or be connected to other tensors depending on the calculation at hand.
See Fig.~\ref{fig: mps}(a) shows how an MPS is drawn in diagrammatic notation.
For a comprehensive introduction to MPS, see e.g. Refs.~\cite{schollwock2011the-density-matrix,verstraete2008matrix}.

\subsubsection{Matrix product operators}
The idea behind MPS can easily be extended to describe operators/matrices \cite{verstraete2004matrix,zwolak2004mixed-state,pirvu2010matrix}.
A matrix product operator or MPO is thus an operator
\beq
     \hat{O} = \sum_{j_{1}=1}^{d} \ldots \sum_{j_{N}=1}^{d} \sum_{i_{1}=1}^{d} \ldots \sum_{i_{N}=1}^{d} O_{j_{1},\dots,j_{N}}^{i_{1},\dots,i_{N}} \ket{j_{1} \dots j_{N}}\bra{i_{1} \dots i_{N}}
\eeq
whose coefficients can be written as as a product of matrices, 
\beq
     \hat{O} =  \sum_{j_{1}=1}^{d} \ldots \sum_{j_{N}=1}^{d} \sum_{i_{1}=1}^{d} \ldots \sum_{i_{N}=1}^{d}
     \Tr \left[ B^{(1)}_{j_{1}, i_{1}} \cdots B^{(N)}_{j_{N}, i_{N}} \right]
     \ket{j_{1} \dots j_{N}}\bra{i_{1} \dots i_{N}},
     \label{mpo}
\eeq 
where $B^{(k)}_{j_{k}, i_{k}}$ are $D_{0} \times D_{0}$ matrices. 
When $j_{k}$ and $i_{k}$ are unspecified, each $B^{(k)}_{j_{k}, i_{k}}$ is a rank-4 tensor.
MPOs are particularly useful for describing local operators, such as local one-dimensional Hamiltonians, which take take an exact MPO representation with some small finite $D_{O}$.
For the case of the OQEM, the Hamiltonian $\hat{H}$ can be written exactly with bond dimension $D_{O} = 3$.
We show an MPO in diagrammatic notation in Fig.~\ref{fig: mps}(b).
It is easy to see why the diagrammatic notation is convenient if one considers the inner product $\braket{\psi | \hat{O} | \psi}$.
The amalgamation of Eqs.~\eqref{mps} and \eqref{mpo} becomes rather complicated.
However, it is simple in the diagrammatic notation, see Fig.~\ref{fig: mps}(c).
Notice that in this case, each of the physical dimensions connect to another tensor, as the inner product is a scalar quantity.

The MPO ansatz can also be used to represent the density matrix corresponding to a mixed state.
Indeed, we are able to use \er{mpo} with $\hat{O} = \rho$.
However, it is important to note that there is no way to enforce positivity in the general MPO ansatz.
Instead, one could use a {\em purification ansatz} \cite{verstraete2004matrix} such that positivity is an innate property.
However, this comes with additional computational costs, and is more restricted than an MPO.
On the other hand, many MPO methods are known to approximately preserve positivity.
More explicitly, if we start with an MPO describing exactly a valid physical state (i.e. positive) and apply physical operations, positivity will be preserved up to the truncation errors introduced by our approximation.
For these reasons, we proceed with an MPO representation.

The density matrix as an MPO is shown in Figure~\ref{fig: mps}(d).
For reasons which will be clear later, it will be useful to vectorise the operator space via the Choi-Jamiołkowski isomorphism. That is, we map 
$\ket{j_{k}}\bra{i_{k}} \to \ket{j_{k}}\ket{i_{k}}$.
For some density matrix $\rho$, we write the vectorised form as $\ket{\rho}$.
This transformation is illustrated in Fig.~\ref{fig: mps}(d); note that we still draw both of the physical legs, but just align them to indicate that we are considering a density matrix as a vectorised operator.
Additionally, when an MPO/MPS is used to describe a density matrix, we will describe its bond dimension by $\chi$ to distinguish it from pure states.
The Lindbladian superoperator $\mathcal{L}$ 
becomes a linear operator in this vectorised space, and is shown in MPO representation in Fig.~\ref{fig: mps}(e).
Under the isomorphism, the Lindbladian takes the form
\beq
     \mathcal{L} \to \hat{\mathcal{L}} = -i(\hat{H} \otimes \hat{I} - \hat{I} \otimes \hat{H}^{T}) +
     \sum_{\alpha=+,-}\sum_{j=1}^{N} \hat{J}_{\alpha, j} \otimes \hat{J}_{\alpha, j}^{*} - \frac{1}{2} \hat{J}_{\alpha, j}^{\dagger}\hat{J}_{\alpha, j} \otimes \hat{I} - \frac{1}{2}\hat{I} \otimes \hat{J}_{\alpha, j}^{T}\hat{J}^{*}_{\alpha, j},
\eeq
where $\hat{I}$ is the identity matrix.
For the OQEM, $\hat{\mathcal{L}}$ can be implemented using an MPO with bond dimension $D_{O} = 5$.
As was done for pure states, we can now write the ``expectation'' of $\rho$ with the Lindbladian superoperator $\Tr\left(\rho\mathcal{L}[\rho]\right) = \Tr \left( \mathcal{L}^{\dagger}[\rho] \rho\right) = \braket{\rho | \hat{\mathcal{L}} | \rho}$ as a TN, see Fig.~\ref{fig: mps}(f).

\subsection{Quantum jump dynamics with matrix product states}

One popular approach for simulating open quantum systems is the quantum jump method~\cite{daley2014qtraj}.
In this method, one uses an unravelling of the Lindbladian (which governs the dynamics of mixed states) onto a {\em stochastic} pure state dynamics.
While the choice of unravelling is not unique, a popular choice is the following.
The pure state evolves via two components: a deterministic evolution under an effective (non-unitary) Hamiltonian $\hat{H}_{\rm eff} = \hat{H} - \frac{i}{2}\sum_{\alpha, j} \hat{J}^{\dagger}_{\alpha, j}\hat{J}_{\alpha, j}$,
\beq
     \ket{\psi_{t+\Delta t}} = \frac{e^{-i\Delta t\hat{H}_{\rm eff}} \ket{\psi_t}}{|| e^{-i\Delta t\hat{H}_{\rm eff}} \ket{\psi_t} ||},
\eeq
and a stochastic evolution which instantaneously applies the jump operators $\hat{J}_{\alpha, j}$ to the wavefunction,
\beq
     \ket{\psi_{t}} \to \frac{\hat{J}_{\alpha, j} \ket{\psi_{t}}} {|| \hat{J}_{\alpha, j} \ket{\psi_{t}} ||}
\eeq
with jump rates
\beq
     w_{\alpha, j}(\ket{\psi_{t}}) = \frac{\braket{\psi_{t} | \hat{J}_{\alpha, j}^{\dagger} \hat{J}_{\alpha, j} | \psi_{t}}}{\braket{\psi_{t} | \psi_{t}}}.
     \label{jump_rate}
\eeq
When averaged over all possible trajectories, the unravelled quantum jump dynamics will give back the evolution of the mixed state under the Lindbladian.

The quantum jump dynamics can be simulated using Quantum Jump Monte Carlo (QJMC).
While there are various strategies for implementing QJMC, we will consider one where time is discretised.
That is, time increments with small time steps $\Delta t \ll 1$.
For each time step, we conduct the following steps:
\begin{enumerate}
     \item Evolve the wavefunction using the effective Hamiltonian, $\ket{{\psi_{t}}'} = e^{-\Delta t \hat{H}_{\rm eff}} \ket{\psi_{t}}$, without normalisation.
     \item Measure the probability that a quantum jump occurs, $P = 1 - \braket{{\psi_{t}}' | {\psi_{t}}'}$.
     \item Uniformly draw a random number $r \in [0, 1]$. If $r > P$, then jump to step 6. Otherwise, proceed to the next step.
     \item Calculate the jump rates $w_{\alpha, j}({\ket{\psi_{t}}'})$, and randomly choose the jump $\alpha, j$ with probabilities 
     \beq
     P_{\alpha, j} = \frac{w_{\alpha, j}({\ket{\psi_{t}}'})}{\sum_{\alpha, j} w_{\alpha, j}({\ket{\psi_{t}}'})}.
     \eeq
     \item Apply the jump operator $\hat{J}_{\alpha, j}$ to the wavefunction, $\ket{{\psi_{t}}'} \to \hat{J}_{\alpha, j} \ket{{\psi_{t}}'}$.
     \item Normalise the wavefunction, $\ket{\psi_{t + \Delta t}} = \ket{{\psi_{t}}'} / || \ket{{\psi_{t}}'} ||$.
\end{enumerate}
Notice that the discretisation of time means that this method is implemented with an error.
In particular, in one time step $\Delta t$, it could be that multiple jumps occurs, while this method only allows for a single jump.
Furthermore, the jumps should be able to occur at any point in continuous time, while this method restricts jumps to occur at integer multiples of $\Delta t$.
Nevertheless, these errors are well controlled by the choice of time step $\Delta t$.

\begin{figure}
     \centering
     \includegraphics[width=0.8\linewidth]{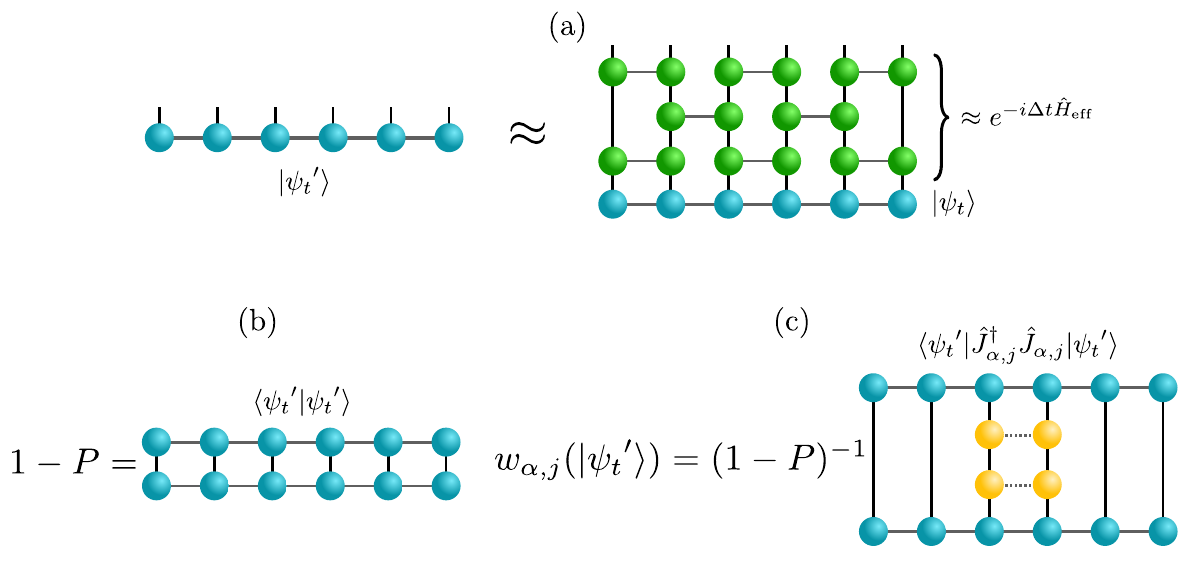}
     \caption{\textbf{Quantum Jump Monte Carlo using matrix product states.} 
     (a) The wavefunction can be evolved under the effective Hamiltonian for some small time step $\Delta t$ using the (second-order) Trotter-Suzuki decomposition.
     The pairs of green tensors are local non-unitary gates.
     (b) After each time-evolution step, we measure the probability of remaining in the state $\ket{\psi_{t}}$, denoted by $\mathcal{N}$.
     (c) The transition rates $w_{\mu}(\ket{\psi_{t}})$ can be calculated as a TN.
     The yellow gates are the two-body jump operators $\hat{J}_{\alpha, j}$ and their adjoints; the dashed lines indicate that the jump operators can be written as two separate tensors with a virtual bond dimension of one.
     }
     \label{fig: qjmc}
 \end{figure}

A common way to implement QJMC is using exact numerics, where the wavefunction is a full-rank vector.
However, this limits the numerics to small system sizes $N \lesssim 20$.
It is simple to generalise the approach for larger system sizes by implementing a TN method \cite{daley2014qtraj,maki2023montecarlo}, where the wavefunction $\ket{\psi_{t}}$ is represented by an MPS.
While, generally speaking, we are unable to exactly implement the time-evolution of the effective Hamiltonian, we can do it approximately using a variety of methods. 
In this paper, we use the second-order Trotter-Suzuki decomposition \cite{trotter1959on-the-product,suzuki1985decomposition}.
The effective Hamiltonian for the OQEM can be written as a sum over two-body operators, $\hat{H}_{\rm eff} = \sum_{j=1}^{N-1} \hat{H}_{j, j+1}$.
We can then write the exponential of the effective Hamiltonian as
\beq
     e^{-\Delta t \hat{H}_{\rm eff}} = \left[ \prod_{j=1}^{\lfloor N/2\rfloor} e^{-i\Delta \hat{H}_{2j-1, 2j}/2} \right]
     \left[ \prod_{j=1}^{\lceil N/2\rceil-1} e^{-i\Delta \hat{H}_{2j, 2j+1}} \right]
     \left[ \prod_{j=1}^{\lfloor N/2\rfloor} e^{-i\Delta \hat{H}_{2j-1, 2j}/2} \right] + \mathcal{O}(\Delta t ^{3}).
     \label{trotter}
\eeq
Since the terms within each product act on different lattice sites, they can be applied simultaneously.
Furthermore, each exponential in the right hand side of \er{trotter} can be calculated exactly, as each only acts on two lattice sites.
\Er{trotter} can be applied directly to the MPS as shown in Fig.~\ref{fig: qjmc}(a). 
The pairs of green tensors represent each exponential in \er{trotter}.
Note that in practice, applying the network of gates to the MPS will result in an MPS with bond dimension $D' \leq 4D$.
However, to prevent the bond dimension growing exponentially in time, we choose to approximate this with another MPS with a smaller bond dimension.
Here, for each gate, we truncate the MPS according to some truncation cutoff threshold $\epsilon = 10^{-12}$, see Ref.~\cite{schollwock2011the-density-matrix} for a definition.
This approach also allows us to dynamically adjust the bond dimension of the MPS as the algorithm proceeds.
The computational cost of calculating the updated MPS is $\mathcal{O}(ND^{3})$.

The probability of a jump occurring can then be measured as a TN calculation, see Fig.~\ref{fig: qjmc}(b).
Contracting the network has computational cost $\mathcal{O}(ND^3)$ (although a canonical representation of the MPS allows for this to be calculated with cost $\mathcal{O}(D^{2})$).
If a jump occurs, then we must also calculate each jump rate \er{jump_rate}.
Each of the $2N$ rates can be calculated as a TN, see Fig.~\ref{fig: qjmc}(c).
Contracting a network such as the one shown in Fig.~\ref{fig: qjmc}(c) has computational cost $\mathcal{O}(ND^{3})$.
It follows that all the jump rates can be calculated with cost $\mathcal{O}(N^{2}D^{3})$.
However, by caching partial contractions, one is able to reduce this to $\mathcal{O}(ND^{3})$ for calculating all the jump rates; see Refs.~\cite{schollwock2011the-density-matrix,causer2021optimal} for details on caching partial contractions.
Note that the choice of unravelling could lead to quantum trajectories with different entanglement properties (and thus a larger bond dimension might be needed). Reference \cite{Vovk2022entanglement} discusses a method to minimise the expected entanglement, and could thus decrease the computational cost needed to simulate the unravelled dynamics.

\subsection{Variational matrix product operators}
In order to find both the spectral gap of the Lindbladian $\mathcal{L}_{s=0}$ and the scaled cumulant generating function (SCGF) of the tilted Lindbladian $\mathcal{L}_{s}$, we employ variational matrix product operators.
The objective is to find eigenstates of $\mathcal{L}_{s}$ with maximal eigenvalue, $\mathcal{L}_{s} [R_{s}] = E R_{s}$ and $\mathcal{L}_{s}^{\dagger} [L_{s}] = E L_{s}$ for the right and left eigenstates respectively.

\subsubsection{Variational matrix product states for self-adjoint operators}
\begin{figure}
     \centering
     \includegraphics[width=0.9\linewidth]{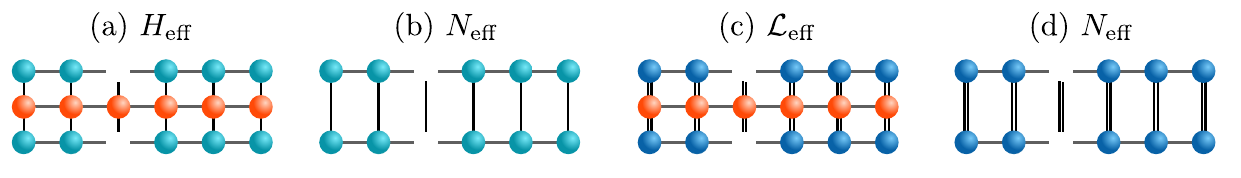}
     \caption{\textbf{Variational matrix product states \& operators.} 
     The (a) effective Hamiltonian $H_{\rm eff}$ and (b) the effective norm $N_{\rm eff}$ used in variational MPS for self-adjoint operators, $\hat{H}$.
     Similarly, we define the (c) effective Lindbladian $\mathcal{L}_{\rm eff}$ and (d) the effective norm $N_{\rm eff}$ for non-self-adjoint $\mathcal{L}$.
     }
     \label{fig: vmps}
 \end{figure}

For a self-adjoint operator, such as a quantum Hamiltonian, $\hat{H}$, this is achieved my maximising (or minimising) the objective function 
\beq
     C(\psi) = \frac{\braket{\psi | \hat{H} | \psi}}{\braket{\psi | \psi}}
     \label{cost_H}
\eeq
for some wavefunction $\psi$.
When $\psi$ is an MPS, maximising (or minimising) \er{cost_H} for some tensor $A_{j}$ in the MPS (while keeping all other tensors fixed) yields the generalised eigenproblem 
\beq
     H_{\rm eff}A_{j} = E N_{\rm eff}A_{j},
     \label{eigen}
\eeq
where the {\em effective Hamiltonian} $H_{\rm eff}$ (not to be confused with the effective Hamiltonian $\hat{H}_{\rm eff}$ in quantum jump dynamics) and the {\em effective norm} $N_{\rm eff}$ are linear operations on $A_{j}$, and are shown as TNs in Figs.~\ref{fig: vmps}(a) and (b).
By using a sparse eigensolver to find the eigenstate with maximal (or minimal) eigenvalue $E$, we find the optimal choice of $A_{j}$.
The procedure is done iteratively for each $j$. Typically, we sweep through each tensor in the MPS from left-to-right and then right-to-left multiple times until convergence in $E$ is met.
The cost of each sweep is $\mathcal{O}(ND^{3})$.

\subsubsection{Variational matrix product operators for non-self-adjoint operators}
For operators which are non-self-adjoint, such as $\hat{L}_{s}$, the variational principle which bounds \er{cost_H} does not apply.
Nevertheless, one can attempt to optimise the MPS using \er{eigen} with a non-Hermitian eigensolver.
Note that here the effective Hamiltonian is replaced with an effective Lindbladian $\mathcal{L}_{\rm eff}$, shown in Fig.~\ref{fig: vmps}(c), and the effective norm is shown in Fig.~\ref{fig: vmps}(d).
While there is no rigorous justification for using this method, in practice, it has been successfully applied to various systems, including quantum master equations \cite{mascarenhas2015}, non-Hermitian quantum Hamiltonians \cite{chan2005density-matrix, Zhang2020skin}, and non-equilibrium stochastic dynamics \cite{helms2019dynamical}.

An alternative approach for estimating the steady state of Lindbladians is to use the above method to minimise $\mathcal{L}^{\dagger} \mathcal{L}$ \cite{cui2015variational}.
This works because the minimal zero-eigenstate of $\mathcal{L}^{\dagger} \mathcal{L}$ corresponds to the zero-eigenstate of $\mathcal{L}$.
However, here we are interested in the spectral gap of $\mathcal{L}_{s}$, and the eigenstates of $\mathcal{L}_{s}$ with the eigenvalue with maximal real component.
These do no have zero eigenvalues, and thus this method cannot be applied.
A recent approach \cite{guo2022variational} adapts the $\mathcal{L}^{\dagger} \mathcal{L}$ approach to minimise $(\mathcal{L} - E)^{\dagger}(\mathcal{L} - E)$, where $E$ is updated using gradient descent (note that this was applied in the context of non-Hermitian Hamiltonians).
We tried this approach here for the tilted generators, and while we had some success, we found that the naive approach of updating the MPS using \er{eigen} with a non-Hermitian eigensolver to be more consistent.

For estimating the spectral gap of Lindbladian, one needs to find the eigenvalue with the second-largest real component.
Given that we know the steady state $\ket{\rho^{\rm ss}}$ of $\hat{\mathcal{L}}$, we can shift the zero-eigenvalue of $\mathcal{L}$ such that the eigenvalue with the second-largest real component now becomes the eigenvalue with largest real component,
$\hat{\mathcal{L}}' = \hat{\mathcal{L}} - \lambda \ket{\rho^{\rm ss}}\bra{\hat{I}}$ for sufficiently large $\lambda > 0$, where $\bra{\hat{I}}$ is the identity matrix in vectorised form.
We can then proceed using the above approach with the shifted Lindbladian.

\subsubsection{Optimisation routine}
We use the non-Hermitian approach described above in the following way to acquire the both the spectral gap of the Lindbladian (for the right eigenvector and projecting out the stationary state), and the leading eigenvectors for the tilted Lindbladian.
We initialise the MPS as some random state with bond dimension $\chi = 1$, and run the non-Hermitian vMPO algorithm. 
Every $100$ sweeps, we measure the expectation with respect to the (tilted) Lindbladian,
\begin{align}
     E_{L} &= \braket{L_{s} | \mathcal{L}_{s}^{\dagger} | L_{s}},
     \label{El}
     \\
     E_{R} &= \braket{R_{s} | \mathcal{L}_{s} | R_{s}},
     \label{Er}
\end{align}
for the left and right eigenvector respectively.
We also measure the ``variance'' of each quantity,
\begin{align}
     \delta E_{L}^{2} &= \braket{L_{s} | \mathcal{L}_{s}\mathcal{L}_{s}^{\dagger} | L_{s}} - \braket{L_{s} | \mathcal{L}_{s} | L_{s}} \braket{L_{s} | \mathcal{L}_{s}^{\dagger} | L_{s}} ,
     \\
     \delta E_{R}^{2} &= \braket{R_{s} | \mathcal{L}_{s}^{\dagger}\mathcal{L}_{s} | R_{s}} - \braket{R_{s} | \mathcal{L}_{s} | R_{s}} \braket{R_{s} | \mathcal{L}_{s}^{\dagger} | R_{s}}.
\end{align}
If the relative change in both the expectation and its variance from the last 100 sweeps is greater than $10^{-3}$, then we perform $100$ more sweeps. This is repeated until this is no longer the case, at which we increase bond dimension ($\chi \to 2\chi$) and proceed.
The ``variance'' indicates how good an approximation the MPS is of an eigenstate of $\hat{\mathcal{L}}_{s}$. 
If $\sqrt{\delta E_{L/R}^{2}} / |E_{L/R}| < \epsilon$ for some bond dimension $\chi$, then the method has converged to the given accuracy, $\epsilon$, and we terminate the optimisation.
We use $\epsilon = 10^{-2}$ for finding the Lindbladian gap, and $\epsilon = 10^{-3}$ for finding the leading eigenvectors of the tilted Lindbladian.
Note that it is not guaranteed that we will find convergence for any $\chi \leq \chi_{\rm max}$, where $\chi_{\rm max} = 1024$ is the maximal bond dimension we allow the algorithm to reach.
In this instance, the quantity $\sqrt{\delta E_{L/R}^{2}} / |E_{L/R}|$ acts as an error metric for the value of $E_{L/R}$.

\subsection{Convergence of the Lindbladian gap}

\begin{figure}
     \centering
     \includegraphics[width=\linewidth]{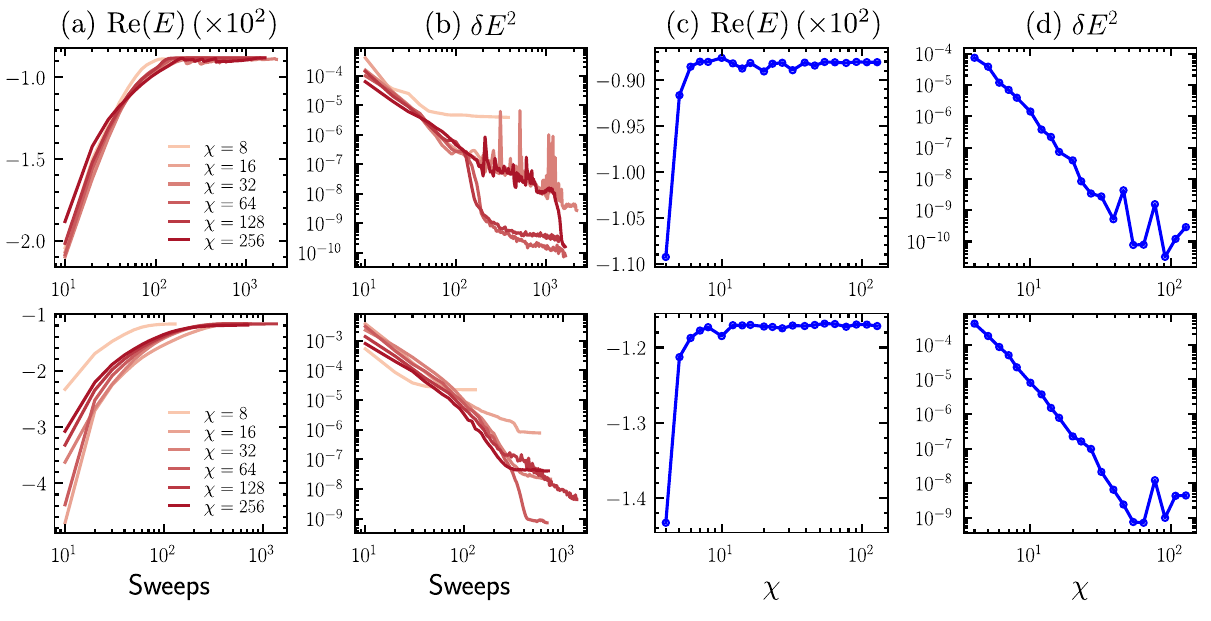}
     \caption{\textbf{Variational estimation of the Lindbladian gap.} 
     Data which demonstrates the convergence of the variational approach for estimating the Lindbladian gap for $N = 30$, $\kappa = 1.0$ and $\gamma = 0.5$.
     The top row is for $\Omega = 0.25$, and the bottom row is for $\Omega = 0.5$.
     (a) The convergence of the expectation of the Lindbladian as a function of sweeps for various bond dimensions, $\chi$.
     (b) The same but for the variance of the expectation.
     (c) The converged expectation of the Lindbladian as a function of of bond dimension $\chi$.
     (d) The same but for the variance of the expectation.
     }
     \label{fig: gap_convergence}
 \end{figure}

In order to probe the relaxation timescales of the OQEM, we used the variational approach described above to estimate the Lindbladian gap.
That is, we estimate the eigenvalue of $\mathcal{L}_{s=0}$ with second-largest real component, and take the negative of its real component to estimate the spectral gap. 
In this subsection, we justify our approach by showing data which demonstrates convergence of the algorithm.
In Fig.~\ref{fig: gap_convergence}, we show data for $N = 30$, $\kappa = 1.0$ and $\gamma = 0.5$.
The top row is for $\Omega = 0.25$, and the bottom row is for $\Omega = 0.5$.

Column (a) of Fig.~\ref{fig: gap_convergence} shows the real part of the expectation $E = \braket{R_{s} | \mathcal{L} | R_{s}}$ as a function of the number of sweeps done in the variational algorithm for various bond dimensions $\chi$.
Notice that in both instances, the algorithm converges to approximately the same value for each bond dimension.
The variance of this expectation $\delta E^{2}$ is shown in column (b). 
We observe that the variance tends to decrease with the number of sweeps.
Note that, compared to vMPS for self-adjoint operators, the optimisation algorithm is less stable, with the variance occasionally spiking.
However, on average, we observe that performing more sweeps increases the accuracy of the MPO approximation.

We also show the real component of $E$ after convergence as a function of bond dimension $\chi$ in column (c).
It is apparent that the observed value of $E$ stabilises as $\chi$ is increased. 
Similarly, in column (d) we show the variance, which steadily decreases with bond dimension.

\subsection{Convergence of dynamical large deviations}

\begin{figure}
     \centering
     \includegraphics[width=\linewidth]{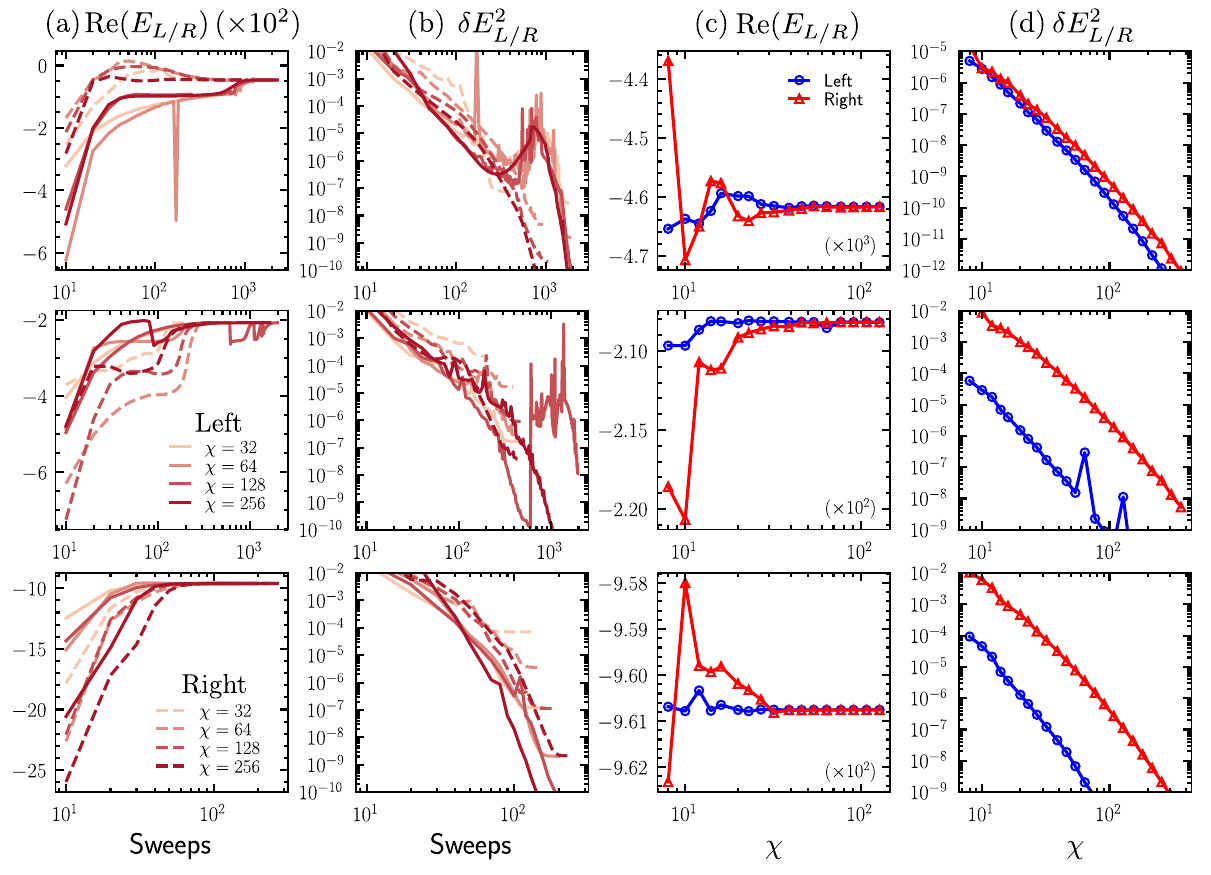}
     \caption{\textbf{Variational estimation of dynamical large deviations.} 
     Data which demonstrates the convergence of the variational approach for estimating the dynamical large deviations for $N = 30$, $\kappa = 1.0$ and $\gamma = 0.0$ and $\Omega = 1.0$.
     The top, middle, and bottom rows are for $s = 10^{-3}, 10^{-2}, 10^{-1}$ respectively.
     (a) The convergence of the expectation of the Lindbladian as a function of sweeps for various bond dimensions, $\chi$, and both the left and right eigenstates.
     (b) The same but for the variance of the expectation.
     (c) The converged expectation of the Lindbladian as a function of bond dimension $\chi$, for both the left (solid lines) and right  (dashed lines) eigenstates. The top row is multiplied by $10^{3}$, and the middle and bottom rows are multiplied by $10^{2}$.
     (d) The same but for the variance of the expectation.
     }
     \label{fig: lds_convergence}
 \end{figure}

The dynamical large deviation statistics can be calculated from the eigenvalue of the tilted generator $\mathcal{L}_{s}$ with largest real component.
In the main text, we used the variational approach described above to estimate both the left and right eigenvectors.
Here, we demonstrate that the variational approach converges with a sufficiently large number of sweeps and bond dimension.
Figure~\ref{fig: lds_convergence} demonstrates this for $N = 30$, $\kappa = 1.0$, $\gamma = 0.0$, $\Omega = 1.0$.
The top, middle, and bottom rows are $s = 10^{-3}, 10^{-2}$ and $10^{-1}$ respectively.

We first show the real component of the expectation of the tilted Lindbladian, Eqs.~\eqref{El} and \eqref{Er}, as a function of the number of sweeps for various bond dimensions $\chi$, see column (a) of Fig.~\ref{fig: lds_convergence}.
The solid lines show this for $E_{L}$, and the dashed lines for $E_{R}$.
Notice that in all instances, each curve converges to approximately the same value, albeit with some instabilities in some instances.
The variance of the same quantity is shown in column (b).
As the algorithm proceeds, the variance steadily decreases. 
At some points, the variance increases, but then resumes its descent. 
These points could be indicative of the algorithm escaping some local minima.

In column (c) of Fig.~\ref{fig: lds_convergence}, we show the convergence of the real component of $E_{L/R}$ with bond dimension $\chi$.  
Notice that the values of $E_{L}$ and $E_{R}$ are in agreement for large $\chi$.
Column (d) shows the variance in $E_{L/R}$.
In each instance, the variance steadily decreases with increased bond dimension.

\subsection{Simulating rare trajectories using matrix product states}

\begin{figure}
     \centering
     \includegraphics[width=0.6\linewidth]{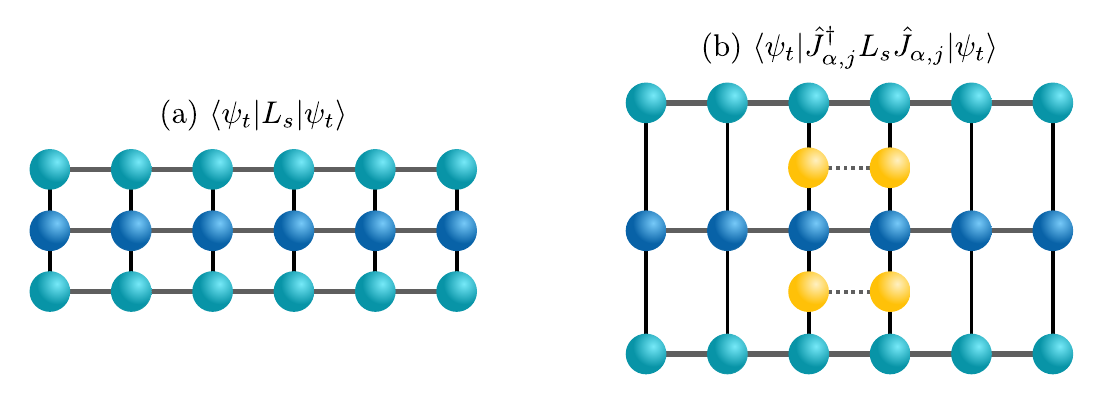}
     \caption{\textbf{Estimation of the transition rates for the Doob dynamics.} 
     The inner products needed to calculate the Doob transition rates as TNs.
     (a) $\braket{\psi_{t} | L_{s} | \psi_{t}}$ and
     (b) $\braket{\psi_{t} | \hat{J}_{\alpha, j}^{\dagger} L_{s} \hat{J}_{\alpha, j} | \psi_{t}}$.
     }
     \label{fig: doob}
 \end{figure}

We now explain how one can use the MPO estimation of the leading left eigenmatrix of $\mathcal{L}$ to implement an approximate dynamics which allows us to directly sample rare fluctuations of the quantum jump dynamics (the so-called Doob dynamics).
From Ref.~\cite{carollo2021large}, an auxiliary process for the full counting statistics can be implemented using the leading eigenmatrix in the following way.
We use the quantum jump dynamics explained earlier in this supplemental material, with the same effective Hamiltonian.
However, the jump rates are instead given by 
\beq
    w^{s}_{\alpha, j}(\ket{\psi_{t}}) = e^{-s}\frac{\braket{\psi_{t} | \hat{J}^{\dagger}_{\alpha, j} L_{s} \hat{J}_{\alpha, j} | \psi_{t}}}{\braket{\psi_{t} | L_{s} | \psi_{t}}} ,
    \label{doob_rates}
\eeq
where $L_{s}$ is our MPO approximation to the left leading eigenmatrix of $\mathcal{L}$.
Both inner products in the fraction are easily calculated as TNs, see Fig.~\ref{fig: doob}.
Calculating all the jump rates can be done with computational cost $\mathcal{O}(ND^{3}\chi + ND^{2}\chi^{2})$, where $D$ is the bond dimension of the MPS wavefunction $\ket{\psi_{t}}$, and $\chi$ is the bond dimension of the MPO $L_{s}$.
Note that in the formulation of the QJMC for the unbiased dynamics (at $s = 0$), the probability of a jump was given by the norm of the time-evolved wavefunction.
This is because the evolution under the effective Hamiltonian encodes the exponential of the time-integrated escape rate (the sum of all the jump rates, integrated over time).
Since the jump rates are altered for $s \neq 0$, this will no longer be the case.
Instead, we measure all the transition rates \er{doob_rates} after each time step.
The probability for a jump is then given by $P \approx 1 - e^{-\Delta t R^{s}(\ket{\psi_{t}})}$, where $R^{s}(\ket{\psi_{t}}) = \sum_{\alpha, j} w^{s}_{\alpha, j}(\ket{\psi_{t}})$.
Note that this introduces an additional source of error, since the escape rate is assumed constant over the time step $\Delta t$.
However, for small $\Delta t$, this error is well-controlled.

\begin{thebibliography}{105}%
\makeatletter
\providecommand \@ifxundefined [1]{%
 \@ifx{#1\undefined}
}%
\providecommand \@ifnum [1]{%
 \ifnum #1\expandafter \@firstoftwo
 \else \expandafter \@secondoftwo
 \fi
}%
\providecommand \@ifx [1]{%
 \ifx #1\expandafter \@firstoftwo
 \else \expandafter \@secondoftwo
 \fi
}%
\providecommand \natexlab [1]{#1}%
\providecommand \enquote  [1]{``#1''}%
\providecommand \bibnamefont  [1]{#1}%
\providecommand \bibfnamefont [1]{#1}%
\providecommand \citenamefont [1]{#1}%
\providecommand \href@noop [0]{\@secondoftwo}%
\providecommand \href [0]{\begingroup \@sanitize@url \@href}%
\providecommand \@href[1]{\@@startlink{#1}\@@href}%
\providecommand \@@href[1]{\endgroup#1\@@endlink}%
\providecommand \@sanitize@url [0]{\catcode `\\12\catcode `\$12\catcode
  `\&12\catcode `\#12\catcode `\^12\catcode `\_12\catcode `\%12\relax}%
\providecommand \@@startlink[1]{}%
\providecommand \@@endlink[0]{}%
\providecommand \url  [0]{\begingroup\@sanitize@url \@url }%
\providecommand \@url [1]{\endgroup\@href {#1}{\urlprefix }}%
\providecommand \urlprefix  [0]{URL }%
\providecommand \Eprint [0]{\href }%
\providecommand \doibase [0]{https://doi.org/}%
\providecommand \selectlanguage [0]{\@gobble}%
\providecommand \bibinfo  [0]{\@secondoftwo}%
\providecommand \bibfield  [0]{\@secondoftwo}%
\providecommand \translation [1]{[#1]}%
\providecommand \BibitemOpen [0]{}%
\providecommand \bibitemStop [0]{}%
\providecommand \bibitemNoStop [0]{.\EOS\space}%
\providecommand \EOS [0]{\spacefactor3000\relax}%
\providecommand \BibitemShut  [1]{\csname bibitem#1\endcsname}%
\let\auto@bib@innerbib\@empty
\bibitem [{\citenamefont {Fredrickson}\ and\ \citenamefont
  {Andersen}(1984)}]{fredrickson1984kinetic}%
  \BibitemOpen
  \bibfield  {author} {\bibinfo {author} {\bibfnamefont {G.~H.}\ \bibnamefont
  {Fredrickson}}\ and\ \bibinfo {author} {\bibfnamefont {H.~C.}\ \bibnamefont
  {Andersen}},\ }\href {https://doi.org/10.1103/PhysRevLett.53.1244} {\bibfield
   {journal} {\bibinfo  {journal} {Phys. Rev. Lett.}\ }\textbf {\bibinfo
  {volume} {53}},\ \bibinfo {pages} {1244} (\bibinfo {year}
  {1984})}\BibitemShut {NoStop}%
\bibitem [{\citenamefont {Palmer}\ \emph {et~al.}(1984)\citenamefont {Palmer},
  \citenamefont {Stein}, \citenamefont {Abrahams},\ and\ \citenamefont
  {Anderson}}]{palmer1984models}%
  \BibitemOpen
  \bibfield  {author} {\bibinfo {author} {\bibfnamefont {R.~G.}\ \bibnamefont
  {Palmer}}, \bibinfo {author} {\bibfnamefont {D.~L.}\ \bibnamefont {Stein}},
  \bibinfo {author} {\bibfnamefont {E.}~\bibnamefont {Abrahams}},\ and\
  \bibinfo {author} {\bibfnamefont {P.~W.}\ \bibnamefont {Anderson}},\ }\href
  {https://doi.org/10.1103/PhysRevLett.53.958} {\bibfield  {journal} {\bibinfo
  {journal} {Phys. Rev. Lett.}\ }\textbf {\bibinfo {volume} {53}},\ \bibinfo
  {pages} {958} (\bibinfo {year} {1984})}\BibitemShut {NoStop}%
\bibitem [{\citenamefont {J{\"a}ckle}\ and\ \citenamefont
  {Eisinger}(1991)}]{jackle1991a-hierarchically}%
  \BibitemOpen
  \bibfield  {author} {\bibinfo {author} {\bibfnamefont {J.}~\bibnamefont
  {J{\"a}ckle}}\ and\ \bibinfo {author} {\bibfnamefont {S.}~\bibnamefont
  {Eisinger}},\ }\href {https://doi.org/10.1007/BF01453764} {\bibfield
  {journal} {\bibinfo  {journal} {Z. Phys. B Condens. Matter}\ }\textbf
  {\bibinfo {volume} {84}},\ \bibinfo {pages} {115} (\bibinfo {year}
  {1991})}\BibitemShut {NoStop}%
\bibitem [{\citenamefont {Kob}\ and\ \citenamefont
  {Andersen}(1993)}]{kob1993kinetic}%
  \BibitemOpen
  \bibfield  {author} {\bibinfo {author} {\bibfnamefont {W.}~\bibnamefont
  {Kob}}\ and\ \bibinfo {author} {\bibfnamefont {H.~C.}\ \bibnamefont
  {Andersen}},\ }\href {https://doi.org/10.1103/PhysRevE.48.4364} {\bibfield
  {journal} {\bibinfo  {journal} {Phys. Rev. E}\ }\textbf {\bibinfo {volume}
  {48}},\ \bibinfo {pages} {4364} (\bibinfo {year} {1993})}\BibitemShut
  {NoStop}%
\bibitem [{\citenamefont {Ritort}\ and\ \citenamefont
  {Sollich}(2003)}]{ritort2003glassy}%
  \BibitemOpen
  \bibfield  {author} {\bibinfo {author} {\bibfnamefont {F.}~\bibnamefont
  {Ritort}}\ and\ \bibinfo {author} {\bibfnamefont {P.}~\bibnamefont
  {Sollich}},\ }\href {https://doi.org/10.1080/0001873031000093582} {\bibfield
  {journal} {\bibinfo  {journal} {Adv. Phys.}\ }\textbf {\bibinfo {volume}
  {52}},\ \bibinfo {pages} {219} (\bibinfo {year} {2003})}\BibitemShut
  {NoStop}%
\bibitem [{\citenamefont {Chandler}\ and\ \citenamefont
  {Garrahan}(2010)}]{chandler2010dynamics}%
  \BibitemOpen
  \bibfield  {author} {\bibinfo {author} {\bibfnamefont {D.}~\bibnamefont
  {Chandler}}\ and\ \bibinfo {author} {\bibfnamefont {J.~P.}\ \bibnamefont
  {Garrahan}},\ }\href {https://doi.org/10.1146/annurev.physchem.040808.090405}
  {\bibfield  {journal} {\bibinfo  {journal} {Annu. Rev. Phys. Chem.}\ }\textbf
  {\bibinfo {volume} {61}},\ \bibinfo {pages} {191} (\bibinfo {year}
  {2010})}\BibitemShut {NoStop}%
\bibitem [{\citenamefont {Garrahan}\ \emph {et~al.}(2011)\citenamefont
  {Garrahan}, \citenamefont {Sollich},\ and\ \citenamefont
  {Toninelli}}]{garrahan2011kinetically}%
  \BibitemOpen
  \bibfield  {author} {\bibinfo {author} {\bibfnamefont {J.~P.}\ \bibnamefont
  {Garrahan}}, \bibinfo {author} {\bibfnamefont {P.}~\bibnamefont {Sollich}},\
  and\ \bibinfo {author} {\bibfnamefont {C.}~\bibnamefont {Toninelli}},\ }in\
  \href@noop {} {\emph {\bibinfo {booktitle} {Dynamical Heterogeneities in
  Glasses, Colloids, and Granular Media}}},\ \bibinfo {series and number}
  {International Series of Monographs on Physics},\ \bibinfo {editor} {edited
  by\ \bibinfo {editor} {\bibfnamefont {L.}~\bibnamefont {Berthier}}, \bibinfo
  {editor} {\bibfnamefont {G.}~\bibnamefont {Biroli}}, \bibinfo {editor}
  {\bibfnamefont {J.-P.}\ \bibnamefont {Bouchaud}}, \bibinfo {editor}
  {\bibfnamefont {L.}~\bibnamefont {Cipelletti}},\ and\ \bibinfo {editor}
  {\bibfnamefont {W.}~\bibnamefont {van Saarloos}}}\ (\bibinfo  {publisher}
  {Oxford University Press},\ \bibinfo {address} {Oxford, UK},\ \bibinfo {year}
  {2011})\ Chap.~\bibinfo {chapter} {10}, pp.\ \bibinfo {pages}
  {341--366}\BibitemShut {NoStop}%
\bibitem [{\citenamefont {Speck}(2019)}]{speck2019dynamic}%
  \BibitemOpen
  \bibfield  {author} {\bibinfo {author} {\bibfnamefont {T.}~\bibnamefont
  {Speck}},\ }\href {https://doi.org/10.1088/1742-5468/ab2ace} {\bibfield
  {journal} {\bibinfo  {journal} {J. Stat. Mech.: Theory Exp.}\ }\textbf
  {\bibinfo {volume} {2019}}\bibinfo  {number} { (8)},\ \bibinfo {pages}
  {084015}}\BibitemShut {NoStop}%
\bibitem [{\citenamefont {Hasyim}\ and\ \citenamefont
  {Mandadapu}(2023)}]{hasyim2023emergent}%
  \BibitemOpen
\bibfield  {number} {  }\bibfield  {author} {\bibinfo {author} {\bibfnamefont
  {M.~R.}\ \bibnamefont {Hasyim}}\ and\ \bibinfo {author} {\bibfnamefont
  {K.~K.}\ \bibnamefont {Mandadapu}},\ }\Eprint
  {https://arxiv.org/abs/2310.06584} {arXiv:2310.06584}  (\bibinfo {year}
  {2023})\BibitemShut {NoStop}%
\bibitem [{\citenamefont {Fendley}\ \emph {et~al.}(2004)\citenamefont
  {Fendley}, \citenamefont {Sengupta},\ and\ \citenamefont
  {Sachdev}}]{fendley2004competing}%
  \BibitemOpen
  \bibfield  {author} {\bibinfo {author} {\bibfnamefont {P.}~\bibnamefont
  {Fendley}}, \bibinfo {author} {\bibfnamefont {K.}~\bibnamefont {Sengupta}},\
  and\ \bibinfo {author} {\bibfnamefont {S.}~\bibnamefont {Sachdev}},\ }\href
  {https://doi.org/10.1103/PhysRevB.69.075106} {\bibfield  {journal} {\bibinfo
  {journal} {Phys. Rev. B}\ }\textbf {\bibinfo {volume} {69}},\ \bibinfo
  {pages} {075106} (\bibinfo {year} {2004})}\BibitemShut {NoStop}%
\bibitem [{\citenamefont {Lesanovsky}(2011)}]{lesanovsky2011many-body}%
  \BibitemOpen
  \bibfield  {author} {\bibinfo {author} {\bibfnamefont {I.}~\bibnamefont
  {Lesanovsky}},\ }\href {https://doi.org/10.1103/PhysRevLett.106.025301}
  {\bibfield  {journal} {\bibinfo  {journal} {Phys. Rev. Lett.}\ }\textbf
  {\bibinfo {volume} {106}},\ \bibinfo {pages} {025301} (\bibinfo {year}
  {2011})}\BibitemShut {NoStop}%
\bibitem [{\citenamefont {Turner}\ \emph {et~al.}(2018)\citenamefont {Turner},
  \citenamefont {Michailidis}, \citenamefont {Abanin}, \citenamefont {Serbyn},\
  and\ \citenamefont {Papi{\'{c}}}}]{turner2018weak}%
  \BibitemOpen
  \bibfield  {author} {\bibinfo {author} {\bibfnamefont {C.~J.}\ \bibnamefont
  {Turner}}, \bibinfo {author} {\bibfnamefont {A.~A.}\ \bibnamefont
  {Michailidis}}, \bibinfo {author} {\bibfnamefont {D.~A.}\ \bibnamefont
  {Abanin}}, \bibinfo {author} {\bibfnamefont {M.}~\bibnamefont {Serbyn}},\
  and\ \bibinfo {author} {\bibfnamefont {Z.}~\bibnamefont {Papi{\'{c}}}},\
  }\href {https://doi.org/10.1038/s41567-018-0137-5} {\bibfield  {journal}
  {\bibinfo  {journal} {Nat. Phys.}\ }\textbf {\bibinfo {volume} {14}},\
  \bibinfo {pages} {745} (\bibinfo {year} {2018})}\BibitemShut {NoStop}%
\bibitem [{\citenamefont {Ates}\ \emph {et~al.}(2007)\citenamefont {Ates},
  \citenamefont {Pohl}, \citenamefont {Pattard},\ and\ \citenamefont
  {Rost}}]{ates2007antiblockade}%
  \BibitemOpen
  \bibfield  {author} {\bibinfo {author} {\bibfnamefont {C.}~\bibnamefont
  {Ates}}, \bibinfo {author} {\bibfnamefont {T.}~\bibnamefont {Pohl}}, \bibinfo
  {author} {\bibfnamefont {T.}~\bibnamefont {Pattard}},\ and\ \bibinfo {author}
  {\bibfnamefont {J.~M.}\ \bibnamefont {Rost}},\ }\href
  {https://doi.org/10.1103/PhysRevLett.98.023002} {\bibfield  {journal}
  {\bibinfo  {journal} {Phys. Rev. Lett.}\ }\textbf {\bibinfo {volume} {98}},\
  \bibinfo {pages} {023002} (\bibinfo {year} {2007})}\BibitemShut {NoStop}%
\bibitem [{\citenamefont {Amthor}\ \emph {et~al.}(2010)\citenamefont {Amthor},
  \citenamefont {Giese}, \citenamefont {Hofmann},\ and\ \citenamefont
  {Weidem\"uller}}]{amthor2010evidence}%
  \BibitemOpen
  \bibfield  {author} {\bibinfo {author} {\bibfnamefont {T.}~\bibnamefont
  {Amthor}}, \bibinfo {author} {\bibfnamefont {C.}~\bibnamefont {Giese}},
  \bibinfo {author} {\bibfnamefont {C.~S.}\ \bibnamefont {Hofmann}},\ and\
  \bibinfo {author} {\bibfnamefont {M.}~\bibnamefont {Weidem\"uller}},\ }\href
  {https://doi.org/10.1103/PhysRevLett.104.013001} {\bibfield  {journal}
  {\bibinfo  {journal} {Phys. Rev. Lett.}\ }\textbf {\bibinfo {volume} {104}},\
  \bibinfo {pages} {013001} (\bibinfo {year} {2010})}\BibitemShut {NoStop}%
\bibitem [{\citenamefont {Lesanovsky}\ and\ \citenamefont
  {Garrahan}(2014)}]{lesanovsky2014out-of-equilibrium}%
  \BibitemOpen
  \bibfield  {author} {\bibinfo {author} {\bibfnamefont {I.}~\bibnamefont
  {Lesanovsky}}\ and\ \bibinfo {author} {\bibfnamefont {J.~P.}\ \bibnamefont
  {Garrahan}},\ }\href {https://doi.org/10.1103/PhysRevA.90.011603} {\bibfield
  {journal} {\bibinfo  {journal} {Phys. Rev. A}\ }\textbf {\bibinfo {volume}
  {90}},\ \bibinfo {pages} {011603} (\bibinfo {year} {2014})}\BibitemShut
  {NoStop}%
\bibitem [{\citenamefont {Hoening}\ \emph {et~al.}(2014)\citenamefont
  {Hoening}, \citenamefont {Abdussalam}, \citenamefont {Fleischhauer},\ and\
  \citenamefont {Pohl}}]{hoening2014antiferromagnetic}%
  \BibitemOpen
  \bibfield  {author} {\bibinfo {author} {\bibfnamefont {M.}~\bibnamefont
  {Hoening}}, \bibinfo {author} {\bibfnamefont {W.}~\bibnamefont {Abdussalam}},
  \bibinfo {author} {\bibfnamefont {M.}~\bibnamefont {Fleischhauer}},\ and\
  \bibinfo {author} {\bibfnamefont {T.}~\bibnamefont {Pohl}},\ }\href
  {https://doi.org/10.1103/PhysRevA.90.021603} {\bibfield  {journal} {\bibinfo
  {journal} {Phys. Rev. A}\ }\textbf {\bibinfo {volume} {90}},\ \bibinfo
  {pages} {021603} (\bibinfo {year} {2014})}\BibitemShut {NoStop}%
\bibitem [{\citenamefont {Valado}\ \emph {et~al.}(2016)\citenamefont {Valado},
  \citenamefont {Simonelli}, \citenamefont {Hoogerland}, \citenamefont
  {Lesanovsky}, \citenamefont {Garrahan}, \citenamefont {Arimondo},
  \citenamefont {Ciampini},\ and\ \citenamefont
  {Morsch}}]{valado2016experimental}%
  \BibitemOpen
  \bibfield  {author} {\bibinfo {author} {\bibfnamefont {M.~M.}\ \bibnamefont
  {Valado}}, \bibinfo {author} {\bibfnamefont {C.}~\bibnamefont {Simonelli}},
  \bibinfo {author} {\bibfnamefont {M.~D.}\ \bibnamefont {Hoogerland}},
  \bibinfo {author} {\bibfnamefont {I.}~\bibnamefont {Lesanovsky}}, \bibinfo
  {author} {\bibfnamefont {J.~P.}\ \bibnamefont {Garrahan}}, \bibinfo {author}
  {\bibfnamefont {E.}~\bibnamefont {Arimondo}}, \bibinfo {author}
  {\bibfnamefont {D.}~\bibnamefont {Ciampini}},\ and\ \bibinfo {author}
  {\bibfnamefont {O.}~\bibnamefont {Morsch}},\ }\href
  {https://doi.org/10.1103/PhysRevA.93.040701} {\bibfield  {journal} {\bibinfo
  {journal} {Phys. Rev. A}\ }\textbf {\bibinfo {volume} {93}},\ \bibinfo
  {pages} {040701} (\bibinfo {year} {2016})}\BibitemShut {NoStop}%
\bibitem [{\citenamefont {Ostmann}\ \emph {et~al.}(2019)\citenamefont
  {Ostmann}, \citenamefont {Marcuzzi}, \citenamefont {Garrahan},\ and\
  \citenamefont {Lesanovsky}}]{ostmann2019localization}%
  \BibitemOpen
  \bibfield  {author} {\bibinfo {author} {\bibfnamefont {M.}~\bibnamefont
  {Ostmann}}, \bibinfo {author} {\bibfnamefont {M.}~\bibnamefont {Marcuzzi}},
  \bibinfo {author} {\bibfnamefont {J.~P.}\ \bibnamefont {Garrahan}},\ and\
  \bibinfo {author} {\bibfnamefont {I.}~\bibnamefont {Lesanovsky}},\ }\href
  {https://doi.org/10.1103/PhysRevA.99.060101} {\bibfield  {journal} {\bibinfo
  {journal} {Phys. Rev. A}\ }\textbf {\bibinfo {volume} {99}},\ \bibinfo
  {pages} {060101} (\bibinfo {year} {2019})}\BibitemShut {NoStop}%
\bibitem [{\citenamefont {Causer}\ \emph {et~al.}(2020)\citenamefont {Causer},
  \citenamefont {Lesanovsky}, \citenamefont {Ba\~nuls},\ and\ \citenamefont
  {Garrahan}}]{causer2020dynamics}%
  \BibitemOpen
  \bibfield  {author} {\bibinfo {author} {\bibfnamefont {L.}~\bibnamefont
  {Causer}}, \bibinfo {author} {\bibfnamefont {I.}~\bibnamefont {Lesanovsky}},
  \bibinfo {author} {\bibfnamefont {M.~C.}\ \bibnamefont {Ba\~nuls}},\ and\
  \bibinfo {author} {\bibfnamefont {J.~P.}\ \bibnamefont {Garrahan}},\ }\href
  {https://doi.org/10.1103/PhysRevE.102.052132} {\bibfield  {journal} {\bibinfo
   {journal} {Phys. Rev. E}\ }\textbf {\bibinfo {volume} {102}},\ \bibinfo
  {pages} {052132} (\bibinfo {year} {2020})}\BibitemShut {NoStop}%
\bibitem [{\citenamefont {van Horssen}\ \emph {et~al.}(2015)\citenamefont {van
  Horssen}, \citenamefont {Levi},\ and\ \citenamefont
  {Garrahan}}]{horssen2015dynamics}%
  \BibitemOpen
  \bibfield  {author} {\bibinfo {author} {\bibfnamefont {M.}~\bibnamefont {van
  Horssen}}, \bibinfo {author} {\bibfnamefont {E.}~\bibnamefont {Levi}},\ and\
  \bibinfo {author} {\bibfnamefont {J.~P.}\ \bibnamefont {Garrahan}},\ }\href
  {https://doi.org/10.1103/PhysRevB.92.100305} {\bibfield  {journal} {\bibinfo
  {journal} {Phys. Rev. B}\ }\textbf {\bibinfo {volume} {92}},\ \bibinfo
  {pages} {100305} (\bibinfo {year} {2015})}\BibitemShut {NoStop}%
\bibitem [{\citenamefont {Lan}\ \emph {et~al.}(2018)\citenamefont {Lan},
  \citenamefont {van Horssen}, \citenamefont {Powell},\ and\ \citenamefont
  {Garrahan}}]{lan2018quantum}%
  \BibitemOpen
  \bibfield  {author} {\bibinfo {author} {\bibfnamefont {Z.}~\bibnamefont
  {Lan}}, \bibinfo {author} {\bibfnamefont {M.}~\bibnamefont {van Horssen}},
  \bibinfo {author} {\bibfnamefont {S.}~\bibnamefont {Powell}},\ and\ \bibinfo
  {author} {\bibfnamefont {J.~P.}\ \bibnamefont {Garrahan}},\ }\href
  {https://doi.org/10.1103/PhysRevLett.121.040603} {\bibfield  {journal}
  {\bibinfo  {journal} {Phys. Rev. Lett.}\ }\textbf {\bibinfo {volume} {121}},\
  \bibinfo {pages} {040603} (\bibinfo {year} {2018})}\BibitemShut {NoStop}%
\bibitem [{\citenamefont {Pancotti}\ \emph {et~al.}(2020)\citenamefont
  {Pancotti}, \citenamefont {Giudice}, \citenamefont {Cirac}, \citenamefont
  {Garrahan},\ and\ \citenamefont {Ba\~nuls}}]{pancotti2020quantum}%
  \BibitemOpen
  \bibfield  {author} {\bibinfo {author} {\bibfnamefont {N.}~\bibnamefont
  {Pancotti}}, \bibinfo {author} {\bibfnamefont {G.}~\bibnamefont {Giudice}},
  \bibinfo {author} {\bibfnamefont {J.~I.}\ \bibnamefont {Cirac}}, \bibinfo
  {author} {\bibfnamefont {J.~P.}\ \bibnamefont {Garrahan}},\ and\ \bibinfo
  {author} {\bibfnamefont {M.~C.}\ \bibnamefont {Ba\~nuls}},\ }\href
  {https://doi.org/10.1103/PhysRevX.10.021051} {\bibfield  {journal} {\bibinfo
  {journal} {Phys. Rev. X}\ }\textbf {\bibinfo {volume} {10}},\ \bibinfo
  {pages} {021051} (\bibinfo {year} {2020})}\BibitemShut {NoStop}%
\bibitem [{\citenamefont {Deger}\ \emph
  {et~al.}(2022{\natexlab{a}})\citenamefont {Deger}, \citenamefont {Roy},\ and\
  \citenamefont {Lazarides}}]{deger2022arresting}%
  \BibitemOpen
  \bibfield  {author} {\bibinfo {author} {\bibfnamefont {A.}~\bibnamefont
  {Deger}}, \bibinfo {author} {\bibfnamefont {S.}~\bibnamefont {Roy}},\ and\
  \bibinfo {author} {\bibfnamefont {A.}~\bibnamefont {Lazarides}},\ }\href
  {https://doi.org/10.1103/PhysRevLett.129.160601} {\bibfield  {journal}
  {\bibinfo  {journal} {Phys. Rev. Lett.}\ }\textbf {\bibinfo {volume} {129}},\
  \bibinfo {pages} {160601} (\bibinfo {year} {2022}{\natexlab{a}})}\BibitemShut
  {NoStop}%
\bibitem [{\citenamefont {Deger}\ \emph
  {et~al.}(2022{\natexlab{b}})\citenamefont {Deger}, \citenamefont
  {Lazarides},\ and\ \citenamefont {Roy}}]{deger2022constrained}%
  \BibitemOpen
  \bibfield  {author} {\bibinfo {author} {\bibfnamefont {A.}~\bibnamefont
  {Deger}}, \bibinfo {author} {\bibfnamefont {A.}~\bibnamefont {Lazarides}},\
  and\ \bibinfo {author} {\bibfnamefont {S.}~\bibnamefont {Roy}},\ }\href
  {https://doi.org/10.1103/PhysRevLett.129.190601} {\bibfield  {journal}
  {\bibinfo  {journal} {Phys. Rev. Lett.}\ }\textbf {\bibinfo {volume} {129}},\
  \bibinfo {pages} {190601} (\bibinfo {year} {2022}{\natexlab{b}})}\BibitemShut
  {NoStop}%
\bibitem [{\citenamefont {Valencia-Tortora}\ \emph {et~al.}(2022)\citenamefont
  {Valencia-Tortora}, \citenamefont {Pancotti},\ and\ \citenamefont
  {Marino}}]{valencia-tortora2022kinetically}%
  \BibitemOpen
  \bibfield  {author} {\bibinfo {author} {\bibfnamefont {R.~J.}\ \bibnamefont
  {Valencia-Tortora}}, \bibinfo {author} {\bibfnamefont {N.}~\bibnamefont
  {Pancotti}},\ and\ \bibinfo {author} {\bibfnamefont {J.}~\bibnamefont
  {Marino}},\ }\href {https://doi.org/10.1103/PRXQuantum.3.020346} {\bibfield
  {journal} {\bibinfo  {journal} {PRX Quantum}\ }\textbf {\bibinfo {volume}
  {3}},\ \bibinfo {pages} {020346} (\bibinfo {year} {2022})}\BibitemShut
  {NoStop}%
\bibitem [{\citenamefont {Carollo}\ \emph {et~al.}(2022)\citenamefont
  {Carollo}, \citenamefont {Gnann}, \citenamefont {Perfetto},\ and\
  \citenamefont {Lesanovsky}}]{carollo2022signatures}%
  \BibitemOpen
  \bibfield  {author} {\bibinfo {author} {\bibfnamefont {F.}~\bibnamefont
  {Carollo}}, \bibinfo {author} {\bibfnamefont {M.}~\bibnamefont {Gnann}},
  \bibinfo {author} {\bibfnamefont {G.}~\bibnamefont {Perfetto}},\ and\
  \bibinfo {author} {\bibfnamefont {I.}~\bibnamefont {Lesanovsky}},\ }\href
  {https://doi.org/10.1103/PhysRevB.106.094315} {\bibfield  {journal} {\bibinfo
   {journal} {Phys. Rev. B}\ }\textbf {\bibinfo {volume} {106}},\ \bibinfo
  {pages} {094315} (\bibinfo {year} {2022})}\BibitemShut {NoStop}%
\bibitem [{\citenamefont {Zadnik}\ \emph {et~al.}(2021)\citenamefont {Zadnik},
  \citenamefont {Bidzhiev},\ and\ \citenamefont
  {Fagotti}}]{zadnik2021the-folded}%
  \BibitemOpen
  \bibfield  {author} {\bibinfo {author} {\bibfnamefont {L.}~\bibnamefont
  {Zadnik}}, \bibinfo {author} {\bibfnamefont {K.}~\bibnamefont {Bidzhiev}},\
  and\ \bibinfo {author} {\bibfnamefont {M.}~\bibnamefont {Fagotti}},\ }\href
  {https://doi.org/10.21468/SciPostPhys.10.5.099} {\bibfield  {journal}
  {\bibinfo  {journal} {SciPost Phys.}\ }\textbf {\bibinfo {volume} {10}},\
  \bibinfo {pages} {099} (\bibinfo {year} {2021})}\BibitemShut {NoStop}%
\bibitem [{\citenamefont {Bertini}\ \emph
  {et~al.}(2023{\natexlab{a}})\citenamefont {Bertini}, \citenamefont {Kos},\
  and\ \citenamefont {Prosen}}]{bertini2023localised}%
  \BibitemOpen
  \bibfield  {author} {\bibinfo {author} {\bibfnamefont {B.}~\bibnamefont
  {Bertini}}, \bibinfo {author} {\bibfnamefont {P.}~\bibnamefont {Kos}},\ and\
  \bibinfo {author} {\bibfnamefont {T.}~\bibnamefont {Prosen}},\ }\href@noop {}
  {\bibinfo {title} {Localised dynamics in the floquet quantum east model}}
  (\bibinfo {year} {2023}{\natexlab{a}}),\ \Eprint
  {https://arxiv.org/abs/2306.12467} {arXiv:2306.12467} \BibitemShut {NoStop}%
\bibitem [{\citenamefont {Bertini}\ \emph
  {et~al.}(2023{\natexlab{b}})\citenamefont {Bertini}, \citenamefont {Fazio},
  \citenamefont {Garrahan},\ and\ \citenamefont {Klobas}}]{bertini2023exact}%
  \BibitemOpen
  \bibfield  {author} {\bibinfo {author} {\bibfnamefont {B.}~\bibnamefont
  {Bertini}}, \bibinfo {author} {\bibfnamefont {C.~D.}\ \bibnamefont {Fazio}},
  \bibinfo {author} {\bibfnamefont {J.~P.}\ \bibnamefont {Garrahan}},\ and\
  \bibinfo {author} {\bibfnamefont {K.}~\bibnamefont {Klobas}},\ }\href@noop {}
  {\bibinfo {title} {Exact quench dynamics of the floquet quantum east model at
  the deterministic point}} (\bibinfo {year} {2023}{\natexlab{b}}),\ \Eprint
  {https://arxiv.org/abs/2310.06128} {arXiv:2310.06128} \BibitemShut {NoStop}%
\bibitem [{\citenamefont {Olmos}\ \emph {et~al.}(2012)\citenamefont {Olmos},
  \citenamefont {Lesanovsky},\ and\ \citenamefont
  {Garrahan}}]{olmos2012facilitated}%
  \BibitemOpen
  \bibfield  {author} {\bibinfo {author} {\bibfnamefont {B.}~\bibnamefont
  {Olmos}}, \bibinfo {author} {\bibfnamefont {I.}~\bibnamefont {Lesanovsky}},\
  and\ \bibinfo {author} {\bibfnamefont {J.~P.}\ \bibnamefont {Garrahan}},\
  }\href {https://doi.org/10.1103/PhysRevLett.109.020403} {\bibfield  {journal}
  {\bibinfo  {journal} {Phys. Rev. Lett.}\ }\textbf {\bibinfo {volume} {109}},\
  \bibinfo {pages} {020403} (\bibinfo {year} {2012})}\BibitemShut {NoStop}%
\bibitem [{\citenamefont {Or{\'u}s}(2019)}]{orus2019tensor}%
  \BibitemOpen
  \bibfield  {author} {\bibinfo {author} {\bibfnamefont {R.}~\bibnamefont
  {Or{\'u}s}},\ }\href {https://doi.org/10.1038/s42254-019-0086-7} {\bibfield
  {journal} {\bibinfo  {journal} {Nat. Revs. Phys.}\ }\textbf {\bibinfo
  {volume} {1}},\ \bibinfo {pages} {538} (\bibinfo {year} {2019})}\BibitemShut
  {NoStop}%
\bibitem [{\citenamefont {Ba\~{n}uls}(2023)}]{banuls2023tensor}%
  \BibitemOpen
  \bibfield  {author} {\bibinfo {author} {\bibfnamefont {M.~C.}\ \bibnamefont
  {Ba\~{n}uls}},\ }\href
  {https://doi.org/10.1146/annurev-conmatphys-040721-022705} {\bibfield
  {journal} {\bibinfo  {journal} {Annu. Rev. Condens. Matter Phys.}\ }\textbf
  {\bibinfo {volume} {14}},\ \bibinfo {pages} {173} (\bibinfo {year}
  {2023})}\BibitemShut {NoStop}%
\bibitem [{\citenamefont {Verstraete}\ \emph {et~al.}(2008)\citenamefont
  {Verstraete}, \citenamefont {Murg},\ and\ \citenamefont
  {Cirac}}]{verstraete2008matrix}%
  \BibitemOpen
  \bibfield  {author} {\bibinfo {author} {\bibfnamefont {F.}~\bibnamefont
  {Verstraete}}, \bibinfo {author} {\bibfnamefont {V.}~\bibnamefont {Murg}},\
  and\ \bibinfo {author} {\bibfnamefont {J.}~\bibnamefont {Cirac}},\ }\href
  {https://doi.org/10.1080/14789940801912366} {\bibfield  {journal} {\bibinfo
  {journal} {Adv. Phys}\ }\textbf {\bibinfo {volume} {57}},\ \bibinfo {pages}
  {143} (\bibinfo {year} {2008})}\BibitemShut {NoStop}%
\bibitem [{\citenamefont
  {Schollw{\"o}ck}(2011)}]{schollwock2011the-density-matrix}%
  \BibitemOpen
  \bibfield  {author} {\bibinfo {author} {\bibfnamefont {U.}~\bibnamefont
  {Schollw{\"o}ck}},\ }\href
  {https://doi.org/https://doi.org/10.1016/j.aop.2010.09.012} {\bibfield
  {journal} {\bibinfo  {journal} {Ann. Phys.}\ }\textbf {\bibinfo {volume}
  {326}},\ \bibinfo {pages} {96} (\bibinfo {year} {2011})},\ \bibinfo {note}
  {january 2011 Special Issue}\BibitemShut {NoStop}%
\bibitem [{\citenamefont {Daley}(2014)}]{daley2014qtraj}%
  \BibitemOpen
  \bibfield  {author} {\bibinfo {author} {\bibfnamefont {A.~J.}\ \bibnamefont
  {Daley}},\ }\href {https://doi.org/10.1080/00018732.2014.933502} {\bibfield
  {journal} {\bibinfo  {journal} {Adv. Phys.}\ }\textbf {\bibinfo {volume}
  {63}},\ \bibinfo {pages} {77} (\bibinfo {year} {2014})}\BibitemShut {NoStop}%
\bibitem [{\citenamefont {Cui}\ \emph {et~al.}(2015)\citenamefont {Cui},
  \citenamefont {Cirac},\ and\ \citenamefont {Ba\~nuls}}]{cui2015variational}%
  \BibitemOpen
  \bibfield  {author} {\bibinfo {author} {\bibfnamefont {J.}~\bibnamefont
  {Cui}}, \bibinfo {author} {\bibfnamefont {J.~I.}\ \bibnamefont {Cirac}},\
  and\ \bibinfo {author} {\bibfnamefont {M.~C.}\ \bibnamefont {Ba\~nuls}},\
  }\href {https://doi.org/10.1103/PhysRevLett.114.220601} {\bibfield  {journal}
  {\bibinfo  {journal} {Phys. Rev. Lett.}\ }\textbf {\bibinfo {volume} {114}},\
  \bibinfo {pages} {220601} (\bibinfo {year} {2015})}\BibitemShut {NoStop}%
\bibitem [{\citenamefont {Markland}\ \emph {et~al.}(2011)\citenamefont
  {Markland}, \citenamefont {Morrone}, \citenamefont {Berne}, \citenamefont
  {Miyazaki}, \citenamefont {Rabani},\ and\ \citenamefont
  {Reichman}}]{markland2011quantum}%
  \BibitemOpen
  \bibfield  {author} {\bibinfo {author} {\bibfnamefont {T.~E.}\ \bibnamefont
  {Markland}}, \bibinfo {author} {\bibfnamefont {J.~A.}\ \bibnamefont
  {Morrone}}, \bibinfo {author} {\bibfnamefont {B.~J.}\ \bibnamefont {Berne}},
  \bibinfo {author} {\bibfnamefont {K.}~\bibnamefont {Miyazaki}}, \bibinfo
  {author} {\bibfnamefont {E.}~\bibnamefont {Rabani}},\ and\ \bibinfo {author}
  {\bibfnamefont {D.~R.}\ \bibnamefont {Reichman}},\ }\href@noop {} {\bibfield
  {journal} {\bibinfo  {journal} {Nat. Phys.}\ }\textbf {\bibinfo {volume}
  {7}},\ \bibinfo {pages} {134} (\bibinfo {year} {2011})}\BibitemShut {NoStop}%
\bibitem [{\citenamefont {Touchette}(2009)}]{touchette2009the-large}%
  \BibitemOpen
  \bibfield  {author} {\bibinfo {author} {\bibfnamefont {H.}~\bibnamefont
  {Touchette}},\ }\href
  {https://doi.org/https://doi.org/10.1016/j.physrep.2009.05.002} {\bibfield
  {journal} {\bibinfo  {journal} {Phys. Rep.}\ }\textbf {\bibinfo {volume}
  {478}},\ \bibinfo {pages} {1} (\bibinfo {year} {2009})}\BibitemShut {NoStop}%
\bibitem [{\citenamefont {Garrahan}(2018)}]{garrahan2018aspects}%
  \BibitemOpen
  \bibfield  {author} {\bibinfo {author} {\bibfnamefont {J.~P.}\ \bibnamefont
  {Garrahan}},\ }\href
  {https://doi.org/https://doi.org/10.1016/j.physa.2017.12.149} {\bibfield
  {journal} {\bibinfo  {journal} {Physica A: Stat. Mech. Appl.}\ }\textbf
  {\bibinfo {volume} {504}},\ \bibinfo {pages} {130} (\bibinfo {year}
  {2018})},\ \bibinfo {note} {lecture Notes of the 14th International Summer
  School on Fundamental Problems in Statistical Physics}\BibitemShut {NoStop}%
\bibitem [{\citenamefont {Jack}(2020)}]{jack2020ergodicity}%
  \BibitemOpen
  \bibfield  {author} {\bibinfo {author} {\bibfnamefont {R.~L.}\ \bibnamefont
  {Jack}},\ }\href {https://doi.org/10.1140/epjb/e2020-100605-3} {\bibfield
  {journal} {\bibinfo  {journal} {Eur. Phys. J. B}\ }\textbf {\bibinfo {volume}
  {93}},\ \bibinfo {pages} {74} (\bibinfo {year} {2020})}\BibitemShut {NoStop}%
\bibitem [{\citenamefont {Garrahan}\ \emph {et~al.}(2007)\citenamefont
  {Garrahan}, \citenamefont {Jack}, \citenamefont {Lecomte}, \citenamefont
  {Pitard}, \citenamefont {van Duijvendijk},\ and\ \citenamefont {van
  Wijland}}]{garrahan2007dynamical}%
  \BibitemOpen
  \bibfield  {author} {\bibinfo {author} {\bibfnamefont {J.~P.}\ \bibnamefont
  {Garrahan}}, \bibinfo {author} {\bibfnamefont {R.~L.}\ \bibnamefont {Jack}},
  \bibinfo {author} {\bibfnamefont {V.}~\bibnamefont {Lecomte}}, \bibinfo
  {author} {\bibfnamefont {E.}~\bibnamefont {Pitard}}, \bibinfo {author}
  {\bibfnamefont {K.}~\bibnamefont {van Duijvendijk}},\ and\ \bibinfo {author}
  {\bibfnamefont {F.}~\bibnamefont {van Wijland}},\ }\href
  {https://doi.org/10.1103/PhysRevLett.98.195702} {\bibfield  {journal}
  {\bibinfo  {journal} {Phys. Rev. Lett.}\ }\textbf {\bibinfo {volume} {98}},\
  \bibinfo {pages} {195702} (\bibinfo {year} {2007})}\BibitemShut {NoStop}%
\bibitem [{\citenamefont {Gorissen}\ \emph {et~al.}(2009)\citenamefont
  {Gorissen}, \citenamefont {Hooyberghs},\ and\ \citenamefont
  {Vanderzande}}]{gorissen2009density-matrix}%
  \BibitemOpen
  \bibfield  {author} {\bibinfo {author} {\bibfnamefont {M.}~\bibnamefont
  {Gorissen}}, \bibinfo {author} {\bibfnamefont {J.}~\bibnamefont
  {Hooyberghs}},\ and\ \bibinfo {author} {\bibfnamefont {C.}~\bibnamefont
  {Vanderzande}},\ }\href {https://doi.org/10.1103/PhysRevE.79.020101}
  {\bibfield  {journal} {\bibinfo  {journal} {Phys. Rev. E}\ }\textbf {\bibinfo
  {volume} {79}},\ \bibinfo {pages} {020101} (\bibinfo {year}
  {2009})}\BibitemShut {NoStop}%
\bibitem [{\citenamefont {Gorissen}\ and\ \citenamefont
  {Vanderzande}(2012)}]{gorissen2012current}%
  \BibitemOpen
  \bibfield  {author} {\bibinfo {author} {\bibfnamefont {M.}~\bibnamefont
  {Gorissen}}\ and\ \bibinfo {author} {\bibfnamefont {C.}~\bibnamefont
  {Vanderzande}},\ }\href {https://doi.org/10.1103/PhysRevE.86.051114}
  {\bibfield  {journal} {\bibinfo  {journal} {Phys. Rev. E}\ }\textbf {\bibinfo
  {volume} {86}},\ \bibinfo {pages} {051114} (\bibinfo {year}
  {2012})}\BibitemShut {NoStop}%
\bibitem [{\citenamefont {Gorissen}\ \emph {et~al.}(2012)\citenamefont
  {Gorissen}, \citenamefont {Lazarescu}, \citenamefont {Mallick},\ and\
  \citenamefont {Vanderzande}}]{gorissen2012exact}%
  \BibitemOpen
  \bibfield  {author} {\bibinfo {author} {\bibfnamefont {M.}~\bibnamefont
  {Gorissen}}, \bibinfo {author} {\bibfnamefont {A.}~\bibnamefont {Lazarescu}},
  \bibinfo {author} {\bibfnamefont {K.}~\bibnamefont {Mallick}},\ and\ \bibinfo
  {author} {\bibfnamefont {C.}~\bibnamefont {Vanderzande}},\ }\href
  {https://doi.org/10.1103/PhysRevLett.109.170601} {\bibfield  {journal}
  {\bibinfo  {journal} {Phys. Rev. Lett.}\ }\textbf {\bibinfo {volume} {109}},\
  \bibinfo {pages} {170601} (\bibinfo {year} {2012})}\BibitemShut {NoStop}%
\bibitem [{\citenamefont {Ba\~nuls}\ and\ \citenamefont
  {Garrahan}(2019)}]{banuls2019using}%
  \BibitemOpen
  \bibfield  {author} {\bibinfo {author} {\bibfnamefont {M.~C.}\ \bibnamefont
  {Ba\~nuls}}\ and\ \bibinfo {author} {\bibfnamefont {J.~P.}\ \bibnamefont
  {Garrahan}},\ }\href {https://doi.org/10.1103/PhysRevLett.123.200601}
  {\bibfield  {journal} {\bibinfo  {journal} {Phys. Rev. Lett.}\ }\textbf
  {\bibinfo {volume} {123}},\ \bibinfo {pages} {200601} (\bibinfo {year}
  {2019})}\BibitemShut {NoStop}%
\bibitem [{\citenamefont {Helms}\ \emph {et~al.}(2019)\citenamefont {Helms},
  \citenamefont {Ray},\ and\ \citenamefont {Chan}}]{helms2019dynamical}%
  \BibitemOpen
  \bibfield  {author} {\bibinfo {author} {\bibfnamefont {P.}~\bibnamefont
  {Helms}}, \bibinfo {author} {\bibfnamefont {U.}~\bibnamefont {Ray}},\ and\
  \bibinfo {author} {\bibfnamefont {G.~K.-L.}\ \bibnamefont {Chan}},\ }\href
  {https://doi.org/10.1103/PhysRevE.100.022101} {\bibfield  {journal} {\bibinfo
   {journal} {Phys. Rev. E}\ }\textbf {\bibinfo {volume} {100}},\ \bibinfo
  {pages} {022101} (\bibinfo {year} {2019})}\BibitemShut {NoStop}%
\bibitem [{\citenamefont {Helms}\ and\ \citenamefont
  {Chan}(2020)}]{helms2020dynamical}%
  \BibitemOpen
  \bibfield  {author} {\bibinfo {author} {\bibfnamefont {P.}~\bibnamefont
  {Helms}}\ and\ \bibinfo {author} {\bibfnamefont {G.~K.-L.}\ \bibnamefont
  {Chan}},\ }\href {https://doi.org/10.1103/PhysRevLett.125.140601} {\bibfield
  {journal} {\bibinfo  {journal} {Phys. Rev. Lett.}\ }\textbf {\bibinfo
  {volume} {125}},\ \bibinfo {pages} {140601} (\bibinfo {year}
  {2020})}\BibitemShut {NoStop}%
\bibitem [{\citenamefont {Causer}\ \emph {et~al.}(2021)\citenamefont {Causer},
  \citenamefont {Ba\~nuls},\ and\ \citenamefont
  {Garrahan}}]{causer2021optimal}%
  \BibitemOpen
  \bibfield  {author} {\bibinfo {author} {\bibfnamefont {L.}~\bibnamefont
  {Causer}}, \bibinfo {author} {\bibfnamefont {M.~C.}\ \bibnamefont
  {Ba\~nuls}},\ and\ \bibinfo {author} {\bibfnamefont {J.~P.}\ \bibnamefont
  {Garrahan}},\ }\href {https://doi.org/10.1103/PhysRevE.103.062144} {\bibfield
   {journal} {\bibinfo  {journal} {Phys. Rev. E}\ }\textbf {\bibinfo {volume}
  {103}},\ \bibinfo {pages} {062144} (\bibinfo {year} {2021})}\BibitemShut
  {NoStop}%
\bibitem [{\citenamefont {Causer}\ \emph {et~al.}(2022)\citenamefont {Causer},
  \citenamefont {Ba\~nuls},\ and\ \citenamefont {Garrahan}}]{causer2022finite}%
  \BibitemOpen
  \bibfield  {author} {\bibinfo {author} {\bibfnamefont {L.}~\bibnamefont
  {Causer}}, \bibinfo {author} {\bibfnamefont {M.~C.}\ \bibnamefont
  {Ba\~nuls}},\ and\ \bibinfo {author} {\bibfnamefont {J.~P.}\ \bibnamefont
  {Garrahan}},\ }\href {https://doi.org/10.1103/PhysRevLett.128.090605}
  {\bibfield  {journal} {\bibinfo  {journal} {Phys. Rev. Lett.}\ }\textbf
  {\bibinfo {volume} {128}},\ \bibinfo {pages} {090605} (\bibinfo {year}
  {2022})}\BibitemShut {NoStop}%
\bibitem [{\citenamefont {Gu}\ and\ \citenamefont
  {Zhang}(2022)}]{gu2022tensor-network}%
  \BibitemOpen
  \bibfield  {author} {\bibinfo {author} {\bibfnamefont {J.}~\bibnamefont
  {Gu}}\ and\ \bibinfo {author} {\bibfnamefont {F.}~\bibnamefont {Zhang}},\
  }\href {https://doi.org/10.1088/1367-2630/ac9ed7} {\bibfield  {journal}
  {\bibinfo  {journal} {New J. Phys.}\ }\textbf {\bibinfo {volume} {24}},\
  \bibinfo {pages} {113022} (\bibinfo {year} {2022})}\BibitemShut {NoStop}%
\bibitem [{\citenamefont {Strand}\ \emph {et~al.}(2022)\citenamefont {Strand},
  \citenamefont {Vroylandt},\ and\ \citenamefont {Gingrich}}]{strand2022using}%
  \BibitemOpen
  \bibfield  {author} {\bibinfo {author} {\bibfnamefont {N.~E.}\ \bibnamefont
  {Strand}}, \bibinfo {author} {\bibfnamefont {H.}~\bibnamefont {Vroylandt}},\
  and\ \bibinfo {author} {\bibfnamefont {T.~R.}\ \bibnamefont {Gingrich}},\
  }\href {https://doi.org/10.1063/5.0097332} {\bibfield  {journal} {\bibinfo
  {journal} {J. Chem. Phys}\ }\textbf {\bibinfo {volume} {156}},\ \bibinfo
  {pages} {221103} (\bibinfo {year} {2022})}\BibitemShut {NoStop}%
\bibitem [{\citenamefont {Causer}\ \emph {et~al.}(2023)\citenamefont {Causer},
  \citenamefont {Ba\~nuls},\ and\ \citenamefont
  {Garrahan}}]{causer2023optimal}%
  \BibitemOpen
  \bibfield  {author} {\bibinfo {author} {\bibfnamefont {L.}~\bibnamefont
  {Causer}}, \bibinfo {author} {\bibfnamefont {M.~C.}\ \bibnamefont
  {Ba\~nuls}},\ and\ \bibinfo {author} {\bibfnamefont {J.~P.}\ \bibnamefont
  {Garrahan}},\ }\href {https://doi.org/10.1103/PhysRevLett.130.147401}
  {\bibfield  {journal} {\bibinfo  {journal} {Phys. Rev. Lett.}\ }\textbf
  {\bibinfo {volume} {130}},\ \bibinfo {pages} {147401} (\bibinfo {year}
  {2023})}\BibitemShut {NoStop}%
\bibitem [{\citenamefont {{Lindblad}}(1976)}]{lindblad1976on-the-generators}%
  \BibitemOpen
  \bibfield  {author} {\bibinfo {author} {\bibfnamefont {G.}~\bibnamefont
  {{Lindblad}}},\ }\href {https://doi.org/10.1007/BF01608499} {\bibfield
  {journal} {\bibinfo  {journal} {Commun. Math. Phys.}\ }\textbf {\bibinfo
  {volume} {48}},\ \bibinfo {pages} {119} (\bibinfo {year} {1976})}\BibitemShut
  {NoStop}%
\bibitem [{\citenamefont {Gorini}\ \emph {et~al.}(1976)\citenamefont {Gorini},
  \citenamefont {Kossakowski},\ and\ \citenamefont
  {Sudarshan}}]{gorini1976completely}%
  \BibitemOpen
  \bibfield  {author} {\bibinfo {author} {\bibfnamefont {V.}~\bibnamefont
  {Gorini}}, \bibinfo {author} {\bibfnamefont {A.}~\bibnamefont
  {Kossakowski}},\ and\ \bibinfo {author} {\bibfnamefont {E.~C.~G.}\
  \bibnamefont {Sudarshan}},\ }\href {https://doi.org/10.1063/1.522979}
  {\bibfield  {journal} {\bibinfo  {journal} {J. Math. Phys.}\ }\textbf
  {\bibinfo {volume} {17}},\ \bibinfo {pages} {821} (\bibinfo {year}
  {1976})}\BibitemShut {NoStop}%
\bibitem [{\citenamefont {Gardiner}\ and\ \citenamefont
  {Zoller}(2004)}]{gardiner2004quantum}%
  \BibitemOpen
  \bibfield  {author} {\bibinfo {author} {\bibfnamefont {C.}~\bibnamefont
  {Gardiner}}\ and\ \bibinfo {author} {\bibfnamefont {P.}~\bibnamefont
  {Zoller}},\ }\href@noop {} {\emph {\bibinfo {title} {{Quantum noise}}}}\
  (\bibinfo  {publisher} {Springer},\ \bibinfo {year} {2004})\BibitemShut
  {NoStop}%
\bibitem [{\citenamefont {Rose}\ \emph {et~al.}(2022)\citenamefont {Rose},
  \citenamefont {Macieszczak}, \citenamefont {Lesanovsky},\ and\ \citenamefont
  {Garrahan}}]{rose2022hierarchical}%
  \BibitemOpen
  \bibfield  {author} {\bibinfo {author} {\bibfnamefont {D.~C.}\ \bibnamefont
  {Rose}}, \bibinfo {author} {\bibfnamefont {K.}~\bibnamefont {Macieszczak}},
  \bibinfo {author} {\bibfnamefont {I.}~\bibnamefont {Lesanovsky}},\ and\
  \bibinfo {author} {\bibfnamefont {J.~P.}\ \bibnamefont {Garrahan}},\ }\href
  {https://doi.org/10.1103/PhysRevE.105.044121} {\bibfield  {journal} {\bibinfo
   {journal} {Phys. Rev. E}\ }\textbf {\bibinfo {volume} {105}},\ \bibinfo
  {pages} {044121} (\bibinfo {year} {2022})}\BibitemShut {NoStop}%
\bibitem [{\citenamefont {Or\'us}(2014)}]{orus2014a-practical}%
  \BibitemOpen
  \bibfield  {author} {\bibinfo {author} {\bibfnamefont {R.}~\bibnamefont
  {Or\'us}},\ }\href {https://doi.org/10.1016/j.aop.2014.06.013} {\bibfield
  {journal} {\bibinfo  {journal} {Ann. Phys.}\ }\textbf {\bibinfo {volume}
  {349}},\ \bibinfo {pages} {117} (\bibinfo {year} {2014})},\ \Eprint
  {https://arxiv.org/abs/1306.2164} {arXiv:1306.2164} \BibitemShut {NoStop}%
\bibitem [{\citenamefont {Bridgeman}\ and\ \citenamefont
  {Chubb}(2017)}]{bridgeman2017hand-waving}%
  \BibitemOpen
  \bibfield  {author} {\bibinfo {author} {\bibfnamefont {J.~C.}\ \bibnamefont
  {Bridgeman}}\ and\ \bibinfo {author} {\bibfnamefont {C.~T.}\ \bibnamefont
  {Chubb}},\ }\href {https://doi.org/10.1088/1751-8121/aa6dc3} {\bibfield
  {journal} {\bibinfo  {journal} {J. Phys. A: Math. Theor.}\ }\textbf {\bibinfo
  {volume} {50}},\ \bibinfo {pages} {223001} (\bibinfo {year}
  {2017})}\BibitemShut {NoStop}%
\bibitem [{\citenamefont {Silvi}\ \emph {et~al.}(2019)\citenamefont {Silvi},
  \citenamefont {Tschirsich}, \citenamefont {Gerster}, \citenamefont
  {J{\"u}nemann}, \citenamefont {Jaschke}, \citenamefont {Rizzi},\ and\
  \citenamefont {Montangero}}]{silvi2019the-tensor}%
  \BibitemOpen
  \bibfield  {author} {\bibinfo {author} {\bibfnamefont {P.}~\bibnamefont
  {Silvi}}, \bibinfo {author} {\bibfnamefont {F.}~\bibnamefont {Tschirsich}},
  \bibinfo {author} {\bibfnamefont {M.}~\bibnamefont {Gerster}}, \bibinfo
  {author} {\bibfnamefont {J.}~\bibnamefont {J{\"u}nemann}}, \bibinfo {author}
  {\bibfnamefont {D.}~\bibnamefont {Jaschke}}, \bibinfo {author} {\bibfnamefont
  {M.}~\bibnamefont {Rizzi}},\ and\ \bibinfo {author} {\bibfnamefont
  {S.}~\bibnamefont {Montangero}},\ }\href
  {https://doi.org/10.21468/SciPostPhysLectNotes.8} {\bibfield  {journal}
  {\bibinfo  {journal} {SciPost Phys. Lect. Notes}\ ,\ \bibinfo {pages} {8}}
  (\bibinfo {year} {2019})}\BibitemShut {NoStop}%
\bibitem [{\citenamefont {Ran}\ \emph {et~al.}(2020)\citenamefont {Ran},
  \citenamefont {Tirrito}, \citenamefont {Peng}, \citenamefont {Chen},
  \citenamefont {Tagliacozzo}, \citenamefont {Su},\ and\ \citenamefont
  {Lewenstein}}]{ran2020tensor}%
  \BibitemOpen
  \bibfield  {author} {\bibinfo {author} {\bibfnamefont {S.-J.}\ \bibnamefont
  {Ran}}, \bibinfo {author} {\bibfnamefont {E.}~\bibnamefont {Tirrito}},
  \bibinfo {author} {\bibfnamefont {C.}~\bibnamefont {Peng}}, \bibinfo {author}
  {\bibfnamefont {X.}~\bibnamefont {Chen}}, \bibinfo {author} {\bibfnamefont
  {L.}~\bibnamefont {Tagliacozzo}}, \bibinfo {author} {\bibfnamefont
  {G.}~\bibnamefont {Su}},\ and\ \bibinfo {author} {\bibfnamefont
  {M.}~\bibnamefont {Lewenstein}},\ }\href
  {https://doi.org/10.1007/978-3-030-34489-4} {\emph {\bibinfo {title} {Tensor
  Network Contractions}}},\ Lecture Notes in Physics\ (\bibinfo  {publisher}
  {Springer International Publishing},\ \bibinfo {year} {2020})\BibitemShut
  {NoStop}%
\bibitem [{\citenamefont {Okunishi}\ \emph {et~al.}(2022)\citenamefont
  {Okunishi}, \citenamefont {Nishino},\ and\ \citenamefont
  {Ueda}}]{okunishi2022developments}%
  \BibitemOpen
  \bibfield  {author} {\bibinfo {author} {\bibfnamefont {K.}~\bibnamefont
  {Okunishi}}, \bibinfo {author} {\bibfnamefont {T.}~\bibnamefont {Nishino}},\
  and\ \bibinfo {author} {\bibfnamefont {H.}~\bibnamefont {Ueda}},\ }\href
  {https://doi.org/10.7566/JPSJ.91.062001} {\bibfield  {journal} {\bibinfo
  {journal} {J. Phys. Soc. Jpn}\ }\textbf {\bibinfo {volume} {91}},\ \bibinfo
  {pages} {062001} (\bibinfo {year} {2022})}\BibitemShut {NoStop}%
\bibitem [{\citenamefont {Cirac}\ \emph {et~al.}(2021)\citenamefont {Cirac},
  \citenamefont {P\'erez-Garc\'{\i}a}, \citenamefont {Schuch},\ and\
  \citenamefont {Verstraete}}]{cirac2021matrix}%
  \BibitemOpen
  \bibfield  {author} {\bibinfo {author} {\bibfnamefont {J.~I.}\ \bibnamefont
  {Cirac}}, \bibinfo {author} {\bibfnamefont {D.}~\bibnamefont
  {P\'erez-Garc\'{\i}a}}, \bibinfo {author} {\bibfnamefont {N.}~\bibnamefont
  {Schuch}},\ and\ \bibinfo {author} {\bibfnamefont {F.}~\bibnamefont
  {Verstraete}},\ }\href {https://doi.org/10.1103/RevModPhys.93.045003}
  {\bibfield  {journal} {\bibinfo  {journal} {Rev. Mod. Phys.}\ }\textbf
  {\bibinfo {volume} {93}},\ \bibinfo {pages} {045003} (\bibinfo {year}
  {2021})}\BibitemShut {NoStop}%
\bibitem [{\citenamefont {Verstraete}\ \emph {et~al.}(2004)\citenamefont
  {Verstraete}, \citenamefont {Garc\'{\i}a-Ripoll},\ and\ \citenamefont
  {Cirac}}]{verstraete2004matrix}%
  \BibitemOpen
  \bibfield  {author} {\bibinfo {author} {\bibfnamefont {F.}~\bibnamefont
  {Verstraete}}, \bibinfo {author} {\bibfnamefont {J.~J.}\ \bibnamefont
  {Garc\'{\i}a-Ripoll}},\ and\ \bibinfo {author} {\bibfnamefont {J.~I.}\
  \bibnamefont {Cirac}},\ }\href
  {https://doi.org/10.1103/PhysRevLett.93.207204} {\bibfield  {journal}
  {\bibinfo  {journal} {Phys. Rev. Lett.}\ }\textbf {\bibinfo {volume} {93}},\
  \bibinfo {pages} {207204} (\bibinfo {year} {2004})}\BibitemShut {NoStop}%
\bibitem [{\citenamefont {Zwolak}\ and\ \citenamefont
  {Vidal}(2004)}]{zwolak2004mixed-state}%
  \BibitemOpen
  \bibfield  {author} {\bibinfo {author} {\bibfnamefont {M.}~\bibnamefont
  {Zwolak}}\ and\ \bibinfo {author} {\bibfnamefont {G.}~\bibnamefont {Vidal}},\
  }\href {https://doi.org/10.1103/PhysRevLett.93.207205} {\bibfield  {journal}
  {\bibinfo  {journal} {Phys. Rev. Lett.}\ }\textbf {\bibinfo {volume} {93}},\
  \bibinfo {pages} {207205} (\bibinfo {year} {2004})}\BibitemShut {NoStop}%
\bibitem [{\citenamefont {Pirvu}\ \emph {et~al.}(2010)\citenamefont {Pirvu},
  \citenamefont {Murg}, \citenamefont {Cirac},\ and\ \citenamefont
  {Verstraete}}]{pirvu2010matrix}%
  \BibitemOpen
  \bibfield  {author} {\bibinfo {author} {\bibfnamefont {B.}~\bibnamefont
  {Pirvu}}, \bibinfo {author} {\bibfnamefont {V.}~\bibnamefont {Murg}},
  \bibinfo {author} {\bibfnamefont {J.~I.}\ \bibnamefont {Cirac}},\ and\
  \bibinfo {author} {\bibfnamefont {F.}~\bibnamefont {Verstraete}},\ }\href
  {http://stacks.iop.org/1367-2630/12/i=2/a=025012} {\bibfield  {journal}
  {\bibinfo  {journal} {New J. Phys.}\ }\textbf {\bibinfo {volume} {12}},\
  \bibinfo {pages} {025012} (\bibinfo {year} {2010})}\BibitemShut {NoStop}%
\bibitem [{\citenamefont {Kliesch}\ \emph {et~al.}(2014)\citenamefont
  {Kliesch}, \citenamefont {Gross},\ and\ \citenamefont
  {Eisert}}]{kliesch2014matrix-product}%
  \BibitemOpen
  \bibfield  {author} {\bibinfo {author} {\bibfnamefont {M.}~\bibnamefont
  {Kliesch}}, \bibinfo {author} {\bibfnamefont {D.}~\bibnamefont {Gross}},\
  and\ \bibinfo {author} {\bibfnamefont {J.}~\bibnamefont {Eisert}},\ }\href
  {https://doi.org/10.1103/PhysRevLett.113.160503} {\bibfield  {journal}
  {\bibinfo  {journal} {Phys. Rev. Lett.}\ }\textbf {\bibinfo {volume} {113}},\
  \bibinfo {pages} {160503} (\bibinfo {year} {2014})}\BibitemShut {NoStop}%
\bibitem [{\citenamefont {De~las Cuevas}\ \emph {et~al.}(2016)\citenamefont
  {De~las Cuevas}, \citenamefont {Cubitt}, \citenamefont {Cirac}, \citenamefont
  {Wolf},\ and\ \citenamefont {P{\'e}rez-Garc{\'\i}a}}]{cuevas2016lim}%
  \BibitemOpen
  \bibfield  {author} {\bibinfo {author} {\bibfnamefont {G.}~\bibnamefont
  {De~las Cuevas}}, \bibinfo {author} {\bibfnamefont {T.~S.}\ \bibnamefont
  {Cubitt}}, \bibinfo {author} {\bibfnamefont {J.~I.}\ \bibnamefont {Cirac}},
  \bibinfo {author} {\bibfnamefont {M.~M.}\ \bibnamefont {Wolf}},\ and\
  \bibinfo {author} {\bibfnamefont {D.}~\bibnamefont {P{\'e}rez-Garc{\'\i}a}},\
  }\href {https://doi.org/10.1063/1.4954983} {\bibfield  {journal} {\bibinfo
  {journal} {J. Math. Phys.}\ }\textbf {\bibinfo {volume} {57}},\ \bibinfo
  {pages} {071902} (\bibinfo {year} {2016})}\BibitemShut {NoStop}%
\bibitem [{Note1()}]{Note1}%
  \BibitemOpen
  \bibinfo {note} {An ansatz worth mentioning is the {\protect \em purification
  ansatz}~\cite
  {verstraete2004matrix,cuevas2013purifications,kliesch2014matrix-product}.While
  this preserves positivity, it is computationally more expensive and more
  restricted.}\BibitemShut {Stop}%
\bibitem [{\citenamefont {Finsterh{\"o}lzl}\ \emph {et~al.}(2020)\citenamefont
  {Finsterh{\"o}lzl}, \citenamefont {Katzer}, \citenamefont {Knorr},\ and\
  \citenamefont {Carmele}}]{finsterholzl2020using}%
  \BibitemOpen
  \bibfield  {author} {\bibinfo {author} {\bibfnamefont {R.}~\bibnamefont
  {Finsterh{\"o}lzl}}, \bibinfo {author} {\bibfnamefont {M.}~\bibnamefont
  {Katzer}}, \bibinfo {author} {\bibfnamefont {A.}~\bibnamefont {Knorr}},\ and\
  \bibinfo {author} {\bibfnamefont {A.}~\bibnamefont {Carmele}},\ }\bibfield
  {journal} {\bibinfo  {journal} {Entropy}\ }\textbf {\bibinfo {volume} {22}},\
  \href {https://doi.org/10.3390/e22090984} {10.3390/e22090984} (\bibinfo
  {year} {2020})\BibitemShut {NoStop}%
\bibitem [{\citenamefont {Garrahan}\ and\ \citenamefont
  {Lesanovsky}(2010)}]{garrahan2010thermodynamics}%
  \BibitemOpen
  \bibfield  {author} {\bibinfo {author} {\bibfnamefont {J.~P.}\ \bibnamefont
  {Garrahan}}\ and\ \bibinfo {author} {\bibfnamefont {I.}~\bibnamefont
  {Lesanovsky}},\ }\href {https://doi.org/10.1103/PhysRevLett.104.160601}
  {\bibfield  {journal} {\bibinfo  {journal} {Phys. Rev. Lett.}\ }\textbf
  {\bibinfo {volume} {104}},\ \bibinfo {pages} {160601} (\bibinfo {year}
  {2010})}\BibitemShut {NoStop}%
\bibitem [{sup()}]{supplemental}%
  \BibitemOpen
  \href@noop {} {\bibinfo {title} {{See Supplemental Material at [URL] for more
  details, which includes Refs. [93-104].}}}\BibitemShut {Stop}%
\bibitem [{\citenamefont {Plenio}\ and\ \citenamefont
  {Knight}(1998)}]{plenio1998the-quantum-jump}%
  \BibitemOpen
  \bibfield  {author} {\bibinfo {author} {\bibfnamefont {M.~B.}\ \bibnamefont
  {Plenio}}\ and\ \bibinfo {author} {\bibfnamefont {P.~L.}\ \bibnamefont
  {Knight}},\ }\href {https://doi.org/10.1103/RevModPhys.70.101} {\bibfield
  {journal} {\bibinfo  {journal} {Rev. Mod. Phys.}\ }\textbf {\bibinfo {volume}
  {70}},\ \bibinfo {pages} {101} (\bibinfo {year} {1998})}\BibitemShut
  {NoStop}%
\bibitem [{\citenamefont {Garrahan}\ and\ \citenamefont
  {Chandler}(2002)}]{garrahan2002geometrical}%
  \BibitemOpen
  \bibfield  {author} {\bibinfo {author} {\bibfnamefont {J.~P.}\ \bibnamefont
  {Garrahan}}\ and\ \bibinfo {author} {\bibfnamefont {D.}~\bibnamefont
  {Chandler}},\ }\href {https://doi.org/10.1103/PhysRevLett.89.035704}
  {\bibfield  {journal} {\bibinfo  {journal} {Phys. Rev. Lett.}\ }\textbf
  {\bibinfo {volume} {89}},\ \bibinfo {pages} {035704} (\bibinfo {year}
  {2002})}\BibitemShut {NoStop}%
\bibitem [{\citenamefont {Blondel}(2013)}]{blondel2013front}%
  \BibitemOpen
  \bibfield  {author} {\bibinfo {author} {\bibfnamefont {O.}~\bibnamefont
  {Blondel}},\ }\href
  {https://doi.org/https://doi.org/10.1016/j.spa.2013.04.014} {\bibfield
  {journal} {\bibinfo  {journal} {Stoch. Process. Their Appl.}\ }\textbf
  {\bibinfo {volume} {123}},\ \bibinfo {pages} {3430} (\bibinfo {year}
  {2013})}\BibitemShut {NoStop}%
\bibitem [{\citenamefont {Merolle}\ \emph {et~al.}(2005)\citenamefont
  {Merolle}, \citenamefont {Garrahan},\ and\ \citenamefont
  {Chandler}}]{merolle2005space-time}%
  \BibitemOpen
  \bibfield  {author} {\bibinfo {author} {\bibfnamefont {M.}~\bibnamefont
  {Merolle}}, \bibinfo {author} {\bibfnamefont {J.~P.}\ \bibnamefont
  {Garrahan}},\ and\ \bibinfo {author} {\bibfnamefont {D.}~\bibnamefont
  {Chandler}},\ }\href
  {https://doi.org/https://doi.org/10.1073/pnas.0504820102} {\bibfield
  {journal} {\bibinfo  {journal} {Proc. Natl. Acad. Sci. USA}\ }\textbf
  {\bibinfo {volume} {102}},\ \bibinfo {pages} {10837} (\bibinfo {year}
  {2005})}\BibitemShut {NoStop}%
\bibitem [{\citenamefont {Katira}\ \emph {et~al.}(2016)\citenamefont {Katira},
  \citenamefont {Mandadapu}, \citenamefont {Vaikuntanathan}, \citenamefont
  {Smit},\ and\ \citenamefont {Chandler}}]{katira2016pre-transition}%
  \BibitemOpen
  \bibfield  {author} {\bibinfo {author} {\bibfnamefont {S.}~\bibnamefont
  {Katira}}, \bibinfo {author} {\bibfnamefont {K.~K.}\ \bibnamefont
  {Mandadapu}}, \bibinfo {author} {\bibfnamefont {S.}~\bibnamefont
  {Vaikuntanathan}}, \bibinfo {author} {\bibfnamefont {B.}~\bibnamefont
  {Smit}},\ and\ \bibinfo {author} {\bibfnamefont {D.}~\bibnamefont
  {Chandler}},\ }\href {https://doi.org/10.7554/eLife.13150} {\bibfield
  {journal} {\bibinfo  {journal} {Elife}\ }\textbf {\bibinfo {volume} {5}},\
  \bibinfo {pages} {e13150} (\bibinfo {year} {2016})}\BibitemShut {NoStop}%
\bibitem [{\citenamefont {Klobas}\ \emph {et~al.}(2023)\citenamefont {Klobas},
  \citenamefont {Fazio},\ and\ \citenamefont {Garrahan}}]{klobas2023exact}%
  \BibitemOpen
  \bibfield  {author} {\bibinfo {author} {\bibfnamefont {K.}~\bibnamefont
  {Klobas}}, \bibinfo {author} {\bibfnamefont {C.~D.}\ \bibnamefont {Fazio}},\
  and\ \bibinfo {author} {\bibfnamefont {J.~P.}\ \bibnamefont {Garrahan}},\
  }\href@noop {} {\bibinfo {title} {Exact "hydrophobicity" in deterministic
  circuits: dynamical fluctuations in the floquet-east model}} (\bibinfo {year}
  {2023}),\ \Eprint {https://arxiv.org/abs/2305.07423} {arXiv:2305.07423}
  \BibitemShut {NoStop}%
\bibitem [{Note2()}]{Note2}%
  \BibitemOpen
  \bibinfo {note} {This is different from the timescale to relax from the most
  unfavourable initial state, as in Fig.~\ref {fig: trajectories}, whose
  timescale is related to the ``cutoff phenomenon'' \cite {aldous1986shuffling}
  of Markov chains.}\BibitemShut {Stop}%
\bibitem [{\citenamefont {Cai}\ and\ \citenamefont
  {Barthel}(2013)}]{cai2013algebraic}%
  \BibitemOpen
  \bibfield  {author} {\bibinfo {author} {\bibfnamefont {Z.}~\bibnamefont
  {Cai}}\ and\ \bibinfo {author} {\bibfnamefont {T.}~\bibnamefont {Barthel}},\
  }\href {https://doi.org/10.1103/PhysRevLett.111.150403} {\bibfield  {journal}
  {\bibinfo  {journal} {Phys. Rev. Lett.}\ }\textbf {\bibinfo {volume} {111}},\
  \bibinfo {pages} {150403} (\bibinfo {year} {2013})}\BibitemShut {NoStop}%
\bibitem [{\citenamefont {Znidaric}(2015)}]{znidaric2015relaxation}%
  \BibitemOpen
  \bibfield  {author} {\bibinfo {author} {\bibfnamefont {M.}~\bibnamefont
  {Znidaric}},\ }\href {https://link.aps.org/doi/10.1103/PhysRevE.92.042143}
  {\bibfield  {journal} {\bibinfo  {journal} {Phys. Rev. E}\ }\textbf {\bibinfo
  {volume} {92}},\ \bibinfo {pages} {042143} (\bibinfo {year}
  {2015})}\BibitemShut {NoStop}%
\bibitem [{\citenamefont {Macieszczak}\ \emph {et~al.}(2016)\citenamefont
  {Macieszczak}, \citenamefont {Guta}, \citenamefont {Lesanovsky},\ and\
  \citenamefont {Garrahan}}]{macieszczak2016towards}%
  \BibitemOpen
  \bibfield  {author} {\bibinfo {author} {\bibfnamefont {K.}~\bibnamefont
  {Macieszczak}}, \bibinfo {author} {\bibfnamefont {M.}~\bibnamefont {Guta}},
  \bibinfo {author} {\bibfnamefont {I.}~\bibnamefont {Lesanovsky}},\ and\
  \bibinfo {author} {\bibfnamefont {J.~P.}\ \bibnamefont {Garrahan}},\ }\href
  {https://doi.org/10.1103/PhysRevLett.116.240404} {\bibfield  {journal}
  {\bibinfo  {journal} {Phys. Rev. Lett.}\ }\textbf {\bibinfo {volume} {116}},\
  \bibinfo {pages} {240404} (\bibinfo {year} {2016})}\BibitemShut {NoStop}%
\bibitem [{\citenamefont {Zhou}\ \emph {et~al.}(2022)\citenamefont {Zhou},
  \citenamefont {Wang},\ and\ \citenamefont {Chen}}]{zhou2022exponential}%
  \BibitemOpen
  \bibfield  {author} {\bibinfo {author} {\bibfnamefont {B.}~\bibnamefont
  {Zhou}}, \bibinfo {author} {\bibfnamefont {X.}~\bibnamefont {Wang}},\ and\
  \bibinfo {author} {\bibfnamefont {S.}~\bibnamefont {Chen}},\ }\href
  {https://doi.org/10.1103/PhysRevB.106.064203} {\bibfield  {journal} {\bibinfo
   {journal} {Phys. Rev. B}\ }\textbf {\bibinfo {volume} {106}},\ \bibinfo
  {pages} {064203} (\bibinfo {year} {2022})}\BibitemShut {NoStop}%
\bibitem [{\citenamefont {Lecomte}\ and\ \citenamefont
  {Tailleur}(2007)}]{lecomte2007a-numerical}%
  \BibitemOpen
  \bibfield  {author} {\bibinfo {author} {\bibfnamefont {V.}~\bibnamefont
  {Lecomte}}\ and\ \bibinfo {author} {\bibfnamefont {J.}~\bibnamefont
  {Tailleur}},\ }\href {http://stacks.iop.org/1742-5468/2007/i=03/a=P03004}
  {\bibfield  {journal} {\bibinfo  {journal} {J. Stat. Mech.}\ }\textbf
  {\bibinfo {volume} {2007}},\ \bibinfo {pages} {P03004} (\bibinfo {year}
  {2007})}\BibitemShut {NoStop}%
\bibitem [{\citenamefont {Maes}(2020)}]{maes2020frenesy:}%
  \BibitemOpen
  \bibfield  {author} {\bibinfo {author} {\bibfnamefont {C.}~\bibnamefont
  {Maes}},\ }\href
  {https://doi.org/https://doi.org/10.1016/j.physrep.2020.01.002} {\bibfield
  {journal} {\bibinfo  {journal} {Phys. Rep.}\ }\textbf {\bibinfo {volume}
  {850}},\ \bibinfo {pages} {1} (\bibinfo {year} {2020})}\BibitemShut {NoStop}%
\bibitem [{Note3()}]{Note3}%
  \BibitemOpen
  \bibinfo {note} {For the OQEM this is guaranteed as ${\protect \mathcal L}$
  is gapped and its stationary state is unique.}\BibitemShut {Stop}%
\bibitem [{\citenamefont {Carollo}\ \emph {et~al.}(2018)\citenamefont
  {Carollo}, \citenamefont {Garrahan}, \citenamefont {Lesanovsky},\ and\
  \citenamefont {P\'erez-Espigares}}]{carollo2018making}%
  \BibitemOpen
  \bibfield  {author} {\bibinfo {author} {\bibfnamefont {F.}~\bibnamefont
  {Carollo}}, \bibinfo {author} {\bibfnamefont {J.~P.}\ \bibnamefont
  {Garrahan}}, \bibinfo {author} {\bibfnamefont {I.}~\bibnamefont
  {Lesanovsky}},\ and\ \bibinfo {author} {\bibfnamefont {C.}~\bibnamefont
  {P\'erez-Espigares}},\ }\href {https://doi.org/10.1103/PhysRevA.98.010103}
  {\bibfield  {journal} {\bibinfo  {journal} {Phys. Rev. A}\ }\textbf {\bibinfo
  {volume} {98}},\ \bibinfo {pages} {010103} (\bibinfo {year}
  {2018})}\BibitemShut {NoStop}%
\bibitem [{\citenamefont {Carollo}\ \emph {et~al.}(2019)\citenamefont
  {Carollo}, \citenamefont {Jack},\ and\ \citenamefont
  {Garrahan}}]{carollo2019unraveling}%
  \BibitemOpen
  \bibfield  {author} {\bibinfo {author} {\bibfnamefont {F.}~\bibnamefont
  {Carollo}}, \bibinfo {author} {\bibfnamefont {R.~L.}\ \bibnamefont {Jack}},\
  and\ \bibinfo {author} {\bibfnamefont {J.~P.}\ \bibnamefont {Garrahan}},\
  }\href {https://doi.org/10.1103/PhysRevLett.122.130605} {\bibfield  {journal}
  {\bibinfo  {journal} {Phys. Rev. Lett.}\ }\textbf {\bibinfo {volume} {122}},\
  \bibinfo {pages} {130605} (\bibinfo {year} {2019})}\BibitemShut {NoStop}%
\bibitem [{\citenamefont {Carollo}\ \emph {et~al.}(2021)\citenamefont
  {Carollo}, \citenamefont {Garrahan},\ and\ \citenamefont
  {Jack}}]{carollo2021large}%
  \BibitemOpen
  \bibfield  {author} {\bibinfo {author} {\bibfnamefont {F.}~\bibnamefont
  {Carollo}}, \bibinfo {author} {\bibfnamefont {J.~P.}\ \bibnamefont
  {Garrahan}},\ and\ \bibinfo {author} {\bibfnamefont {R.~L.}\ \bibnamefont
  {Jack}},\ }\href {https://doi.org/10.1007/s10955-021-02799-x} {\bibfield
  {journal} {\bibinfo  {journal} {J. Stat. Phys.}\ }\textbf {\bibinfo {volume}
  {184}},\ \bibinfo {pages} {13} (\bibinfo {year} {2021})}\BibitemShut
  {NoStop}%
\bibitem [{\citenamefont {Borkar}\ \emph {et~al.}(2003)\citenamefont {Borkar},
  \citenamefont {Juneja},\ and\ \citenamefont
  {Kherani}}]{borkar2003peformance}%
  \BibitemOpen
  \bibfield  {author} {\bibinfo {author} {\bibfnamefont {V.~S.}\ \bibnamefont
  {Borkar}}, \bibinfo {author} {\bibfnamefont {S.}~\bibnamefont {Juneja}},\
  and\ \bibinfo {author} {\bibfnamefont {A.~A.}\ \bibnamefont {Kherani}},\
  }\href {https://projecteuclid.org:443/euclid.cis/1119639799} {\bibfield
  {journal} {\bibinfo  {journal} {Commun. Inf. Syst.}\ }\textbf {\bibinfo
  {volume} {3}},\ \bibinfo {pages} {259} (\bibinfo {year} {2003})}\BibitemShut
  {NoStop}%
\bibitem [{\citenamefont {Jack}\ and\ \citenamefont
  {Sollich}(2010)}]{jack2010large}%
  \BibitemOpen
  \bibfield  {author} {\bibinfo {author} {\bibfnamefont {R.~L.}\ \bibnamefont
  {Jack}}\ and\ \bibinfo {author} {\bibfnamefont {P.}~\bibnamefont {Sollich}},\
  }\href {https://doi.org/10.1143/PTPS.184.304} {\bibfield  {journal} {\bibinfo
   {journal} {Prog. Theor. Phys. Supp.}\ }\textbf {\bibinfo {volume} {184}},\
  \bibinfo {pages} {304} (\bibinfo {year} {2010})}\BibitemShut {NoStop}%
\bibitem [{\citenamefont {Chetrite}\ and\ \citenamefont
  {Touchette}(2015)}]{chetrite2015nonequilibrium}%
  \BibitemOpen
  \bibfield  {author} {\bibinfo {author} {\bibfnamefont {R.}~\bibnamefont
  {Chetrite}}\ and\ \bibinfo {author} {\bibfnamefont {H.}~\bibnamefont
  {Touchette}},\ }\href
  {https://doi.org/https://doi.org/10.1007/s00023-014-0375-8} {\bibfield
  {journal} {\bibinfo  {journal} {Ann. Henri Poincar{\'{e}}}\ }\textbf
  {\bibinfo {volume} {16}},\ \bibinfo {pages} {2005} (\bibinfo {year}
  {2015})}\BibitemShut {NoStop}%
\bibitem [{\citenamefont {Garrahan}(2016)}]{garrahan2016classical}%
  \BibitemOpen
  \bibfield  {author} {\bibinfo {author} {\bibfnamefont {J.~P.}\ \bibnamefont
  {Garrahan}},\ }\href {https://doi.org/10.1088/1742-5468/2016/07/073208}
  {\bibfield  {journal} {\bibinfo  {journal} {J. Stat. Mech.: Theory Exp}\
  }\textbf {\bibinfo {volume} {2016}},\ \bibinfo {pages} {073208} (\bibinfo
  {year} {2016})}\BibitemShut {NoStop}%
\bibitem [{\citenamefont {Hastings}(2007)}]{hastings2007an-area}%
  \BibitemOpen
  \bibfield  {author} {\bibinfo {author} {\bibfnamefont {M.~B.}\ \bibnamefont
  {Hastings}},\ }\href {https://doi.org/10.1088/1742-5468/2007/08/P08024}
  {\bibfield  {journal} {\bibinfo  {journal} {J. Stat. Mech.: Theory Exp}\
  }\textbf {\bibinfo {volume} {2007}},\ \bibinfo {pages} {P08024} (\bibinfo
  {year} {2007})}\BibitemShut {NoStop}%
\bibitem [{\citenamefont {Verstraete}\ and\ \citenamefont
  {Cirac}(2006)}]{verstraete2006matrix}%
  \BibitemOpen
  \bibfield  {author} {\bibinfo {author} {\bibfnamefont {F.}~\bibnamefont
  {Verstraete}}\ and\ \bibinfo {author} {\bibfnamefont {J.~I.}\ \bibnamefont
  {Cirac}},\ }\href {https://doi.org/10.1103/PhysRevB.73.094423} {\bibfield
  {journal} {\bibinfo  {journal} {Phys. Rev. B}\ }\textbf {\bibinfo {volume}
  {73}},\ \bibinfo {pages} {094423} (\bibinfo {year} {2006})}\BibitemShut
  {NoStop}%
\bibitem [{\citenamefont {Eisert}\ \emph {et~al.}(2010)\citenamefont {Eisert},
  \citenamefont {Cramer},\ and\ \citenamefont
  {Plenio}}]{eisert2010colloquium:}%
  \BibitemOpen
  \bibfield  {author} {\bibinfo {author} {\bibfnamefont {J.}~\bibnamefont
  {Eisert}}, \bibinfo {author} {\bibfnamefont {M.}~\bibnamefont {Cramer}},\
  and\ \bibinfo {author} {\bibfnamefont {M.~B.}\ \bibnamefont {Plenio}},\
  }\href {https://doi.org/10.1103/RevModPhys.82.277} {\bibfield  {journal}
  {\bibinfo  {journal} {Rev. Mod. Phys.}\ }\textbf {\bibinfo {volume} {82}},\
  \bibinfo {pages} {277} (\bibinfo {year} {2010})}\BibitemShut {NoStop}%
\bibitem [{\citenamefont {Maki}\ \emph {et~al.}(2023)\citenamefont {Maki},
  \citenamefont {Berti}, \citenamefont {Carusotto},\ and\ \citenamefont
  {Biella}}]{maki2023montecarlo}%
  \BibitemOpen
  \bibfield  {author} {\bibinfo {author} {\bibfnamefont {J.~A.}\ \bibnamefont
  {Maki}}, \bibinfo {author} {\bibfnamefont {A.}~\bibnamefont {Berti}},
  \bibinfo {author} {\bibfnamefont {I.}~\bibnamefont {Carusotto}},\ and\
  \bibinfo {author} {\bibfnamefont {A.}~\bibnamefont {Biella}},\ }\href
  {https://doi.org/10.21468/SciPostPhys.15.4.152} {\bibfield  {journal}
  {\bibinfo  {journal} {SciPost Phys.}\ }\textbf {\bibinfo {volume} {15}},\
  \bibinfo {pages} {152} (\bibinfo {year} {2023})}\BibitemShut {NoStop}%
\bibitem [{\citenamefont {Trotter}(1959)}]{trotter1959on-the-product}%
  \BibitemOpen
  \bibfield  {author} {\bibinfo {author} {\bibfnamefont {H.~F.}\ \bibnamefont
  {Trotter}},\ }\href
  {https://www.ams.org/journals/proc/1959-010-04/S0002-9939-1959-0108732-6/}
  {\bibfield  {journal} {\bibinfo  {journal} {Proc. American Math. Soc.}\
  }\textbf {\bibinfo {volume} {10}},\ \bibinfo {pages} {545} (\bibinfo {year}
  {1959})}\BibitemShut {NoStop}%
\bibitem [{\citenamefont {Vovk}\ and\ \citenamefont
  {Pichler}(2022)}]{Vovk2022entanglement}%
  \BibitemOpen
  \bibfield  {author} {\bibinfo {author} {\bibfnamefont {T.}~\bibnamefont
  {Vovk}}\ and\ \bibinfo {author} {\bibfnamefont {H.}~\bibnamefont {Pichler}},\
  }\href {https://doi.org/10.1103/PhysRevLett.128.243601} {\bibfield  {journal}
  {\bibinfo  {journal} {Phys. Rev. Lett.}\ }\textbf {\bibinfo {volume} {128}},\
  \bibinfo {pages} {243601} (\bibinfo {year} {2022})}\BibitemShut {NoStop}%
\bibitem [{\citenamefont {Mascarenhas}\ \emph {et~al.}(2015)\citenamefont
  {Mascarenhas}, \citenamefont {Flayac},\ and\ \citenamefont
  {Savona}}]{mascarenhas2015}%
  \BibitemOpen
  \bibfield  {author} {\bibinfo {author} {\bibfnamefont {E.}~\bibnamefont
  {Mascarenhas}}, \bibinfo {author} {\bibfnamefont {H.}~\bibnamefont
  {Flayac}},\ and\ \bibinfo {author} {\bibfnamefont {V.}~\bibnamefont
  {Savona}},\ }\href {https://doi.org/10.1103/PhysRevA.92.022116} {\bibfield
  {journal} {\bibinfo  {journal} {Phys. Rev. A}\ }\textbf {\bibinfo {volume}
  {92}},\ \bibinfo {pages} {022116} (\bibinfo {year} {2015})}\BibitemShut
  {NoStop}%
\bibitem [{\citenamefont {Chan}\ and\ \citenamefont
  {Van~Voorhis}(2005)}]{chan2005density-matrix}%
  \BibitemOpen
  \bibfield  {author} {\bibinfo {author} {\bibfnamefont {G.~K.-L.}\
  \bibnamefont {Chan}}\ and\ \bibinfo {author} {\bibfnamefont {T.}~\bibnamefont
  {Van~Voorhis}},\ }\href {https://doi.org/10.1063/1.1899124} {\bibfield
  {journal} {\bibinfo  {journal} {J. Chem. Phys.}\ }\textbf {\bibinfo {volume}
  {122}},\ \bibinfo {pages} {204101} (\bibinfo {year} {2005})}\BibitemShut
  {NoStop}%
\bibitem [{\citenamefont {Zhang}\ \emph {et~al.}(2020)\citenamefont {Zhang},
  \citenamefont {Chen}, \citenamefont {Zhang}, \citenamefont {Lang},
  \citenamefont {Li},\ and\ \citenamefont {Zhu}}]{Zhang2020skin}%
  \BibitemOpen
  \bibfield  {author} {\bibinfo {author} {\bibfnamefont {D.-W.}\ \bibnamefont
  {Zhang}}, \bibinfo {author} {\bibfnamefont {Y.-L.}\ \bibnamefont {Chen}},
  \bibinfo {author} {\bibfnamefont {G.-Q.}\ \bibnamefont {Zhang}}, \bibinfo
  {author} {\bibfnamefont {L.-J.}\ \bibnamefont {Lang}}, \bibinfo {author}
  {\bibfnamefont {Z.}~\bibnamefont {Li}},\ and\ \bibinfo {author}
  {\bibfnamefont {S.-L.}\ \bibnamefont {Zhu}},\ }\href
  {https://doi.org/10.1103/PhysRevB.101.235150} {\bibfield  {journal} {\bibinfo
   {journal} {Phys. Rev. B}\ }\textbf {\bibinfo {volume} {101}},\ \bibinfo
  {pages} {235150} (\bibinfo {year} {2020})}\BibitemShut {NoStop}%
\bibitem [{\citenamefont {Guo}\ \emph {et~al.}(2022)\citenamefont {Guo},
  \citenamefont {Xu}, \citenamefont {Li}, \citenamefont {You},\ and\
  \citenamefont {Yang}}]{guo2022variational}%
  \BibitemOpen
  \bibfield  {author} {\bibinfo {author} {\bibfnamefont {Z.}~\bibnamefont
  {Guo}}, \bibinfo {author} {\bibfnamefont {Z.-T.}\ \bibnamefont {Xu}},
  \bibinfo {author} {\bibfnamefont {M.}~\bibnamefont {Li}}, \bibinfo {author}
  {\bibfnamefont {L.}~\bibnamefont {You}},\ and\ \bibinfo {author}
  {\bibfnamefont {S.}~\bibnamefont {Yang}},\ }\href@noop {} {\bibinfo {title}
  {{Variational Matrix Product State Approach for Non-Hermitian System Based on
  a Companion Hermitian Hamiltonian}}} (\bibinfo {year} {2022}),\ \Eprint
  {https://arxiv.org/abs/2210.14858} {arXiv:2210.14858} \BibitemShut {NoStop}%
\bibitem [{\citenamefont {de~las Cuevas}\ \emph {et~al.}(2013)\citenamefont
  {de~las Cuevas}, \citenamefont {Schuch}, \citenamefont
  {{P\'erez}-Garc{\'{\i}}a},\ and\ \citenamefont
  {Cirac}}]{cuevas2013purifications}%
  \BibitemOpen
  \bibfield  {author} {\bibinfo {author} {\bibfnamefont {G.}~\bibnamefont
  {de~las Cuevas}}, \bibinfo {author} {\bibfnamefont {N.}~\bibnamefont
  {Schuch}}, \bibinfo {author} {\bibfnamefont {D.}~\bibnamefont
  {{P\'erez}-Garc{\'{\i}}a}},\ and\ \bibinfo {author} {\bibfnamefont {J.~I.}\
  \bibnamefont {Cirac}},\ }\href
  {http://stacks.iop.org/1367-2630/15/i=12/a=123021} {\bibfield  {journal}
  {\bibinfo  {journal} {New J. Phys.}\ }\textbf {\bibinfo {volume} {15}},\
  \bibinfo {pages} {123021} (\bibinfo {year} {2013})}\BibitemShut {NoStop}%
\bibitem [{\citenamefont {Aldous}\ and\ \citenamefont
  {Diaconis}(1986)}]{aldous1986shuffling}%
  \BibitemOpen
  \bibfield  {author} {\bibinfo {author} {\bibfnamefont {D.}~\bibnamefont
  {Aldous}}\ and\ \bibinfo {author} {\bibfnamefont {P.}~\bibnamefont
  {Diaconis}},\ }\href {https://doi.org/10.2307/2323590} {\bibfield  {journal}
  {\bibinfo  {journal} {Am. Math. Mon.}\ }\textbf {\bibinfo {volume} {93}},\
  \bibinfo {pages} {333} (\bibinfo {year} {1986})}\BibitemShut {NoStop}%
\bibitem [{\citenamefont {Suzuki}(1985)}]{suzuki1985decomposition}%
  \BibitemOpen
  \bibfield  {author} {\bibinfo {author} {\bibfnamefont {M.}~\bibnamefont
  {Suzuki}},\ }\href {https://doi.org/10.1063/1.526596} {\bibfield  {journal}
  {\bibinfo  {journal} {J. Math. Phys.}\ }\textbf {\bibinfo {volume} {26}},\
  \bibinfo {pages} {601} (\bibinfo {year} {1985})}\BibitemShut {NoStop}%
\end{thebibliography}
\end{document}